\documentclass[12pt,preprint]{aastex}
\begin{document}
\title{The B-band Luminosity function of red and blue galaxies\\
 up to $z=3.5$}

\author{E. Giallongo\altaffilmark{1}, S. Salimbeni\altaffilmark{1}, N. Menci\altaffilmark{1}, G. Zamorani\altaffilmark{2},
A. Fontana\altaffilmark{1}, M. Dickinson\altaffilmark{3}, S. Cristiani\altaffilmark{4}, L. Pozzetti\altaffilmark{2}}
\altaffiltext{1}{INAF, Osservatorio Astronomico di Roma, via Frascati 33, I-00040 Monteporzio, Italy} 
\altaffiltext{2}{INAF, Osservatorio Astronomico di Bologna, via Ranzani 1, I-40127 Bologna, Italy} 
\altaffiltext{3}{National Optical Astronomy Observatory, 950 N. Cherry St., Tucson, AZ 85719} 
\altaffiltext{4}{INAF, Osservatorio Astronomico di Trieste, via G.B. Tiepolo 11, I-34131 Trieste, Italy} 

\begin{abstract}

We  have   explored  the   redshift   evolution   of  the   luminosity
function of  red and  blue galaxies  up to  $z=3.5$. This was possible
joining  a  deep I  band  composite   galaxy  sample,  which  includes
the  spectroscopic K20 sample  and the  HDFs  samples,  with the  deep
$H_{AB}=26$  and $K_{AB}=25$ samples derived  from the deep NIR images
of  the Hubble Deep    Fields North   and South,   respectively. About
30\% of    the      sample     has   spectroscopic   redshifts     and
the remaining   fraction   well-calibrated    photometric   redshifts.
This allowed  to  select   and  measure  galaxies in    the rest-frame
blue  magnitude   up   to $z\sim 3$  and to   derive    the   redshift
evolution  of   the    B-band  luminosity    function    of   galaxies
separated  by   their  rest-frame     $U-V$    color   or     specific
(i.e.     per   unit     mass)   star-formation    rate.   The   class
separation    was   derived   from  passive    evolutionary tracks  or
from    their   observed  bimodal  distributions.  Both  distributions
appear   bimodal   at  least   up to     $z\sim  2$  and the locus  of
red/early galaxies is clearly identified  up to these high  redshifts.
Both  luminosity and density  evolutions are  needed      to  describe
the cosmological behaviour  of the   red/early      and      blue/late
populations.  The  density  evolution   is greater  for   the    early
population   with  a  decrease   by   one   order  of   magnitude   at
$z\sim   2-3$    with   respect  to  the value  at  $z\sim  0.4$.  The
luminosity   densities  of    the early    and late type galaxies with
$M_B<-20.6$ appear      to    have    a  bifurcation    at      $z>1$.
Indeed  while star-forming  galaxies    slightly   increase  or   keep
constant their  luminosity      density,     "early"  galaxies
decrease in  their luminosity   density   by   a  factor $\sim
5-6$ from $z\sim  0.4$   to  $z\sim  2.5-3$.  A  comparison  with  one
of   the latest versions  of  the  hierarchical CDM   models  shows  a
broad  agreement   with    the   observed     number  and   luminosity
density evolutions  of  both populations.

\end{abstract}


\section{Introduction}

The redshift   evolution of   the galaxy  luminosity  function  in
different bands   is   a  powerful   probe   of  the   main   physical
processes   governing   the   assembly  and  subsequent   activity  of
galaxies. In  particular, the   analysis of  the luminosity   function
of galaxies having different spectral and/or morphological  properties
can help to disentangle different histories of star-formation.

In the emerging picture, galaxies form hierarchically through  mergers
of pre-existing smaller  galaxies in the  context of Cold  Dark Matter
(CDM) scenario. In this  picture  early type   galaxies, which are  in
general among  the  brightest and massive   galaxies  in the universe,
tend to  be the  result of  relatively late  assembly. This appears in
contrast with  the old  age inferred  from their  red colors   and
with recent observations  by Cimatti  et al.  (2004) and Glazebrook et
al. (2004)  showing   the   presence  of  some  massive  galaxies   at
high redshift.

For this reason  it is very  important to analyze  the abundance as  a
function of redshift of early  type galaxies selected on the  basis of
their morphological appearance or spectral energy distributions (SED).
This is  a difficult  task both  in the  observational and theoretical
context: on the one side we   need  large and  deep  surveys  of   NIR
selected galaxies to sample the rest-frame visual band at least up  to
$z\sim 3$; on  the   other  side  we   need  sophisticated  MonteCarlo
or hydrodynamical codes  for galaxy  formation and  evolution in   the
CDM scenario.

Most of the information on  the statistical analysis of the  red/early
or blue/late type galaxies comes  from local samples (e.g.  Blanton et
al. 2003, Baldry   et al. 2003,  Brinchmann   et al.  2004). In  these
studies a clear bimodality is  appearing in the color distribution  of
galaxies  or in  their specific  (i.e. per  unit mass)  star-formation
rate.

This bimodality separates the locus of the star-forming galaxies  from
the early population (early type spirals, S0, E0)  and thus represents
a useful tool to  study the redshift evolution of  galaxies having
colors consistent  with those of  the early types independently of  any
morphological   selection.  In  the  local   universe  these different
definitions  appear to be consistent with each other (see  e.g.  Schweitzer
and Seitzer 1992).

In  general it  is difficult  to extract  information about  the  star
formation  histories  of  individual early  type  galaxies  since  the
stellar population evolving  passively can become  red in a  time much
shorter  than  the  age  of  the  galaxy. 
A  complementary   approach   is  given   by   the   analysis  of  the
redshift  evolution  of  the luminosity  function  of  galaxies having
different colors or specific star-formation rates.

Recent attempts in  this respect  have  been made on  the  local Sloan
sample e.g. by Hogg  et   al. (2002), Blanton  et  al.  (2003), Baldry
et  al.   (2004) and   at intermediate    redshifts by   Bell et   al.
(2004).  Preliminary   results  show   that   the   rest-frame   $U-V$
distribution  of galaxies   is   bimodal   up to   $z\sim    1$   even
for  the field population. This allows a   separation of  the   galaxy
population     in    red/early     and    blue/late    galaxies.    In
particular,  the   rest-frame  B    band   luminosity    function   of
the   red  galaxies shows   a rather   flat slope   and evolves   only
mildly in qualitatively good agreement with  the $\Lambda$CDM scenario.

  An  attempt  to  follow at  higher  redshifts  the  evolution of
galaxies of different spectral types has  shown  in   particular  that
the   number     density  of    early  type  galaxies   decreases with
increasing   redshift   (Wolf et     al.   2003).  However the authors
used   fixed   spectral    energy  distributions   to    separate  the
galaxy  spectral   types neglecting evolutionary effects due  to   the
passive stellar   evolution with  redshift.  Moreover, the  lack    of
near   infrared   imaging  in   their  sample     has  limited     the
analysis to  $z\lesssim 1$.

The extension of  this information at  high redshifts requires indeed
deep   NIR surveys  to get   an estimate   of the   rest-frame  $U-V$
color. The  J,H band images in   the  HDFN and  the J,K  band   images
in the   HDFS are among  the deepest   NIR images  present in   fields
where   extensive  multicolor optical   imaging is   already available
and  have  been   used  to  derive   accurate  photometric   redshifts
(Dickinson  et  al.   2003, Labbe  et  al. 2003, Fontana  et al. 2003,
Poli et al. 2003).

In this context, we  have already  used in  Poli et  al. (2003,  paper
I)  a  composite sample  of  NIR selected  galaxies  that,    although
relatively  small, has  unique characteristics   in many     respects.
It   is       a   complete   sample  selected   in the rest-frame    B
band  from  low    to high   redshifts,  just thanks to   the deep NIR
HDF south images. It  includes the area  of the K20   survey  (Cimatti
et     al.    2002)    with         a      large    fraction        of
spectroscopic    redshifts at      the    bright   end      and   well
tested       photometric  redshifts    at    the     faint  end.  This
composite sample  has been  used to  derive the  evolution of the global rest-frame
B-band luminosity function  up to  $z\sim  3$.  The  main  result
was  the  presence  of  appreciable luminosity evolution of the bright
end of the LF. Such   a brightening could  be  produced   by  enhanced
star-formation  at   high  redshifts  triggered by   galaxy encounters
in a  CDM scenario  (see Menci et al. 2004,  Nagamine  et  al.  2004).
This   kind   of   evolution   is  dominated  by    the   star-forming
population, which   is abundant   at high redshift.

On the other hand, we have also explored the evolution in mass of  the
sample, exploring in particular the evolution of the mass function  up
to  intermediate  redshifts  and  the mass  density  up  to  $z\sim 3$
(Fontana et   al. 2003,   2004). This  approach allowed us to follow  the
cosmological evolution of galaxies in a way less biased by episodes of
starburst activity. At $z\lesssim  1$ little evolution is  observed in
the galaxy mass function derived from the K20 sample, while  at higher
redshifts a more  pronounced decrease  in  the  stellar mass   density
was   found in  the  HDF-S    up to    $z\sim   3$ in  broad agreement
with   $\Lambda$CDM   models  (see  also  Rudnick   et   al.    2003).
However, the derived   fraction of mass  density with  respect  to the
local    value  at     these high redshifts shows  field-to-field variations
by   about   a   factor    2  if  we  compare   the results
obtained  in    the    HDF-S    field  with those found
in  the HDF-North  by Dickinson   et al. (2003).

To  explore  how  different  the star  formation  history  can  be  in
different  galaxy  populations,  we  derive  in    this    paper   the
number   and  luminosity  density  evolutions   of the  galaxies  as a
function of their  color  or specific (i.e. per unit stellar mass) SFR
(SSFR)     up     to     $z\sim     3$,  and   we    compare       the
density evolution of the red/early or blue/late  populations with that
predicted  by  a MonteCarlo  rendition  of the  Menci  et  al.  (2004)
$\Lambda$CDM model.
In the paper all the magnitudes are in the AB system and an $\Omega_{\Lambda}=0.7$,
$\Omega_M=0.3$, and $H_0=70$ km s$^{-1}$ Mpc$^{-1}$ cosmology was adopted.

\section{The galaxy sample}

Our composite   sample of  galaxies covers,  with  similar   depth,  a
range   spanning    from   the   UV   down     to   the    optical/NIR
wavelengths  and is described in detail in paper I. Here we  summarize
its   general  characteristics  and  main improvements.  The  galaxies
are  selected  in  the  $I$  or  NIR  $J,K$  bands  depending  on  the
redshift interval  as explained  below. A  first   photometric catalog
has   been derived  combining  the   optical    images  in  the  HDF-S
(Casertano et  al.   2000)    with   the   deep ESO/VLT   NIR   images
obtained in   the   framework   of   the     FIRES    project (Labb\'e
et   al. 2003).     Our   infrared    catalog   reaches    formal    5
sigma  limits       in       an   aperture      of    $1.2''$       at
$J_{AB}=25.3$,   $H_{AB}=24$,       and        $K_{AB}=25$.  In the  I
band the  HDF-S  galaxies  have been  selected down  to $I_{AB}=26.2$. In the
HDF North we  used  the same  $K_{AB}<23.5$ catalog   as  in  paper I.
In addition   to  the   sample used   in  paper   I,   we  adopted the
HDF-N   multicolor catalog   of galaxies   selected   in   the  NICMOS
$H$    band     by  Dickinson  et       al.    (2000)      which    is
highly    ($>95$\%)  complete down        to      $H_{AB}<26$       or
$I_{AB}<25.8$, where     the  relatively   bright  limit    in     the
$I$ band   was  required  to ensure  detection  in  the  NIR    bands.
We used    also a     brighter sample   of  galaxies     selected   in
two fields centered   around the   QSO    0055-269  and      in    the
Chandra  Deep    Field  South  (Giacconi  et     al.   2001)       for
a    total   of      $\simeq  68$      $arcmin^2$.    This      sample
was     used   to    select the   targets         for     the       so
called  K20   ($K_{AB}\le 21.9$)   spectroscopic     survey   (Cimatti
et     al.  2002).    The catalogs  in  the  latter  two  fields  have
been  used   to   derive complete     ($>95$\%)   samples         down
to    $I_{AB}=24.5,25$,  respectively.   Detection    in   the     NIR
bands  is ensured   for all the galaxies  in the   composite sample.

To  obtain   the    total   fluxes    of   the    objects   in     the
chosen reference   band,  SExtractor  Kron   magnitudes  (Bertin    \&
Arnouts  1996)  have    been adopted     for  bright     sources.  For
faint   sources  in  ground-based  images  we   estimated     aperture
magnitudes         with aperture  corrections   computed    estimating
the    flux      losses outside    the adopted  aperture with  respect
to  the  Kron  magnitudes  for   relatively    bright  sources.    The
reliability    of     the correction adopted    has been   tested   by
means    of     simulations  as     performed     in   paper I  and in
Poli  et     al. (2001).      No     appreciable systematic     losses
of   the   total  flux    for     the      faintest   sources     were
found  in addition to     the    well     known losses   by     5-10\%
typical   of  the Kron    magnitudes. Moreover, Bouwens et al.  (2004)
have recently compared images obtained with HST/ACS in the Ultra  Deep
Field  with  those obtained  in  the   same area  from the   shallower
ACS-GOODS  survey.  The main  result is  that there  are no   specific
biases due to  surface brightness  dimming in  the evaluation  of  the
typical  galaxy   sizes   and  consequently   on    the   estimate  of
their  total magnitudes   up    to   high    redshifts.   To      cope
with           the    inhomogeneous        quality      of         the
different images, we produced a set of seeing$-$corrected {\em  color}
magnitudes    that      have  been       then   scaled  to       total
magnitudes  in    all         the       bands (see    e.g.    Vanzella
et      al.  2001).       This     ensures that     colors  -     used
to     estimate  the         redshifts      -         are     measured
in      the      same physical      region at     all     wavelengths,
independently    of  the    seeing   FWHM present   in    each  image.
A fraction    of 29\%   of the  sample has  spectroscopic   redshifts,
mainly from    the     K20   survey   (Cimatti   et     al. 2002) and
from the spectroscopic follow up  in the GOODS/Chandra field (Vanzella
et al. 2004), and in the HDF north (e.g. Cohen et al. 2000).       For
all    the      other     galaxies,    reliable photometric  redshifts
have been derived with a homogeneous  technique, described in  Fontana
et al.  (2000) and  in Cimatti  et al    (2002).  The  model    adopts
several     galaxy      ages     and  exponentially declining     star
formation   histories,     and      includes  Lyman absorption    from
the      intergalactic    medium         and    dust absorption   with
different    amounts       up    to      $E(B -V)=1$, using   both SMC
and     Calzetti    extinction curves.       In all    the    samples,
the relative accuracy is $\sigma _{\Delta z}\lesssim 0.05$ where $\Delta z=
(z_{spe}-z_{phot})/(1+z_{spe})$.

Our combined  sample  allows  to compute the  $4400$ {\AA}  rest frame
LF  in  a   continuous way   from  $z\sim  0.4$  up to $z\sim    3.5$.
 The  galaxy luminosity  in  this rest-frame  wavelength  is  not
strongly affected by dust, is directly observed up to high redshift in
the NIR bands and can be easily compared with local  estimates of  the
blue luminosity  function. As   in  paper   I,  we      used   our $I
-$band   selected    sample for   galaxies  at    $z<1$. Indeed,   the
4400 {\AA}  rest  frame   wavelength is  within  or  shorter  than the
I   band,     and   the   LF   includes  all   the      galaxies  with
$m[4400(1+z)]\leq    I_{AB}(lim)$   of  the  various   samples.   This
implies   that     some    galaxies    from  the  original    $I_{AB}$
limited  samples  are     excluded   from  the LF    since        they
have      a    red    spectrum       and consequently   $m[4400(1+z)]$
are  fainter than  our adopted threshold. We   have     in  this   way
extracted     a    complete  subsample     of  galaxies   having     a
continuum   magnitude    threshold   at    4400   {\AA}   in      each
redshift   interval,   independently      of  their    color (assuming
galaxies    in     this        redshift     interval       have    $(R
-I)_{AB}\geq   0$).  The   same  procedure   has   been adopted     at
higher   $z$  using   the $K$   or $H$  bands    selected sample.   In
the $1.3<z<2.5$  and $2.5<z<3.5$     interval,  the  4400 {\AA}   rest
frame wavelength  is    within    or shortward    of    the H   and  K
bands,  respectively.   The          LF     includes         all   the
galaxies         with  $m[4400(1+z)]=m_{AB}\leq   m_{AB}(lim)$,  where
$m_{AB}\equiv    H_{AB}$   or    $K_{AB}$   for     $1.3<z<2.5$    and
$m_{AB}\equiv K_{AB}$  for $2.5<z<3.5$.    Finally, as    in paper  I,
the rest     frame absolute   AB magnitude   $M_B(AB)$  has       been
derived   from     the  theoretical  best    fit  spectral      energy
distribution  for     each  galaxy.   The  same    fit   provides  the
photometric redshift   for each  galaxy.   In  this     way   the    K
-correction    has       been  derived  for      each   galaxy    from
the    interpolation    between     the observed   magnitudes    using
the        best       fit      spectrum.  Uncertainties    in     this
interpolation    produce     on      average  small         ($\lesssim
10\%$)    errors    in     the        rest-frame  luminosities  (Ellis
1997;   Pozzetti   et     al.  2003).   

\section{The evolution as a function of color and star-formation}

To  explore  the  evolution  of  the  galaxy  luminosity  function for
different  spectral   types   we   have  analyzed   first  the   color
distribution   of galaxies    as   a    function   of    redshift.  In
Fig.1 we show the distribution  of  the $U-V$ colors    as  a function
of $z$  for the   two samples,   the I  band selected   at  $z<1$  and
the $H,K$    band    selected   at    $z=1.3 -3.5$.  The    rest-frame
$U-V$ color has   been used  because    it includes the   4000   {\AA}
break  and     so  it   is   most     sensitive  to    the      galaxy
properties  (age,  star    formation,    etc.). Moreover the     $U-V$
color   is   always  sampled by  the   multicolor catalog   of     our
composite  optical-NIR   sample   in     a   large  redshift  interval
up   to   $z\sim   3.5$.  No correction  for dust absorption has  been
applied to the rest  frame $U -V$ color.

Previous   analyses  on   the  luminosity  function   evolution   have
adopted  a  constant  rest-frame   color to separate    red/early-type
galaxies   from   the     blue/star-forming population (e.g. the  CFRS
sample  (Lilly   et  al.   1995)).   However,    when  extending    this
criterion    to     high  redshifts       we      can   not     ignore
evolutionary   effects in the stellar population of high $z$ galaxies.
As a  first  attempt  to include   this effect  we  assumed a scenario
where  the   star  formation     history  of   the  red/early   galaxy
population  is controlled  by passive  stellar evolution  from a  high
formation     redshift.     Thus      we     adopted     the   PLE
(pure  luminosity  evolution)  evolutionary   color   track    of   a
galaxy   whose   local   color   is  similar  to  the  average   color
of a    typical  S0    galaxy as     a criterion    to separate    the
galaxy   populations.    In     this  way   we      can  follow    the
blueing    of    the      early-type population seen    at  increasing
redshifts     assuming     that  only passive  evolution  was   active
in   the  past.

In Fig.1   the "S0   evolutionary  track"   is  shown    as a function
of     $z$.      A     formation     redshift     of      $z=10$,   an
exponential time-scale  $\tau=3$   Gyr  and  solar  metallicity   have
been adopted  to   provide   at    $z=0$   the    typical   color   $U
-V\sim 2.1\pm 0.1$ of  this   class  of  galaxies (Schweizer \&  Seitzer
1992). The latest version of the standard Bruzual   \& Charlot  (2003)
model  has  been    adopted.  This       evolutionary  track naturally
follows    the  evolution in    redshift  of  the  blue  edge  of  the
underdense    regions  occupied  by  red  objects  in     the observed
bimodal  color distribution.  The color  separation increases as   the
redshift decreases  from $U-V\sim  1$  at  $z\sim 2.5$ to $U-V\sim
1.5$ at $z\sim 0.7$ as  shown in Fig.1. Above the cut, the  spread  in
the  colors  can  be represented  by  passive   evolution  with  short
($\tau =0.3,1,3$      Gyr)     exponential      star     formation
timescales with  a  high formation  redshift ($z=10$).

Although the adopted  evolutionary   color cut is somewhat  arbitrary,
we have  verified   that    small   changes      of    the  parameters
do not  change significantly the results  and  conclusions about   the
evolution  of  the red and blue populations outlined in section 5. The
main   constraint is   that the    evolutionary color  tracks   should
provide  at $z\simeq   0$   colors  still   consistent with   that  of
S0  galaxies. There is room    for  example to reduce   the  formation
redshift  from  $z=10$   to   $z=6$.  The  new  tracks  would  produce
a  blue-shift       by  about     0.1    in    the  color cut at  each
redshift    ($U-V\simeq   2$    at   $z\simeq    0$).  The   resulting
luminosity    function  evolutions    of   the     red   and      blue
populations   do   not   change appreciably      and  the      changes
in       the      evolutionary  parameters     of    the    luminosity
functions   are     well    within the  errors. 

It is clear  however that  if  we are interested   in the analysis  of
the galaxy  evolution as  a function  of  the star-formation activity,
the use  of a  color criterion  introduces  some    population  mixing
since it is not possible  to distinguish an early-type  galaxy  from a
dusty starburst using only the $U-V$ color.

A further step in this direction can be obtained with a   spectral fit
to the  overall  SEDs by  means  of the  Bruzual   \&  Charlot  (2003)
spectral synthesis model. Although  some degeneracy still remains,  we
can remove the most obvious starburst galaxies from the locus of early
type galaxies by considering  the  specific (i.e. per unit mass)  star
formation   rate    $\dot    m_*    /m_*$   as      a    function   of
redshift that is plotted in Fig.2. The use of $\dot m_*  /m_*$ in this
context has a number  of  advantages with respect  to using  the  SFR;
for example  it is largely  independent of cosmology  and it  is  less
sensitive  to    the stellar  IMF  (see  e.g.  Guzman  et   al.  1997,
Brinchmann   and    Ellis   2000,  Brinchmann   et    al.  2004).  The
library used   to   fit  the    galaxy  properties  (SFR   and stellar
mass)  and       to derive    photometric redshifts    for   the faint
fraction  of   the sample   is described in   more  detail   in  paper
I  and  in  Fontana et   al. (2004).  We are  aware that the absolute
values      in    the    $\dot    m_*/m_*$ distribution  are   subject
to uncertainties   due   for   example  to  the    estimate    of  the
dust attenuation  which  depends  on  the extinction   curve   adopted
and to  the   methods      adopted   for the mass       estimate.   We
refer in   particular to   the papers   by Fontana    et   al.  (2004)
and  Dickinson  et  al.   (2003)   for  the detailed   description  of
the    methods  used   to  derive    a    reliable  estimate   of  the
stellar  mass  from  the  SED  fitting   procedure to  our  multicolor
galaxy   sample.    We      have  recently   quantified  a  systematic
difference between  the  stellar mass derived  from  the best fit   to
the     SED  and    that  based   on  the   so  called    "maximal age
approach". The  masses derived   from  the  best fit  method  are   in
general a    factor   $\lesssim 2$ lower   (Fontana  et al.       2004).  Any
physical  interpretation  of      the    values produced    in     the
observed   $\dot   m_*/m_*$       distribution should   take      into
account  all   these uncertainties.

To  explore  how  much  the  evolution  of  the  red  galaxies  can be
influenced by the presence  of dusty starburst galaxies, especially  at high
$z$, we have  separated the population  according to the  evolutionary
track of  the specific  star formation  rate $\dot  m_*/m_*$ 
which corresponds to the "S0 color track" already  adopted.
The evolution  of  the  classes separation   appears evident  in
Fig.2. The evolutionary SSFR decreases from 
$\log(\dot m_*/m_*)\sim  -9.4$ to $\log   (\dot m_*/m_*)\sim   -10.3$
 (in units of yr$^{-1}$) from $z\sim 2.5$ to $z\sim 0.7$. Below the cut,
the   increasing  spread    in  $\dot    m_*/m_*$  with
decreasing  redshifts can be represented  by  passive  evolution  with
short     ($\tau   =0.3,1,3$    Gyr)   exponential    star   formation
timescales with  a  high formation  redshift ($z=10$).

The color/SSFR "S0 evolutionary tracks"  will be used in section  5 to
provide  the  separation  of  the  galaxy  populations  at   different
redshifts and to analyze their  luminosity function  evolutions under the
assumption that  the star  formation history  of the  red galaxies  is
mainly subject to stellar passive evolution.

\section{The bimodal color/star-formation distributions as a function of z}

A  different selection  criterion  which   is  independent   of    the
assumed evolutionary  scenario is   based  on the  empirical  analysis
of  the color or   $\dot  m_*/m_*$   distributions. Previous  analyses
at  lower redshifts ($z<1$) by Bell et al. (2004) and Baldry  et   al.
(2004) have shown that  these   distributions  are   bimodal   and  so
identify   two distinct   populations,  the  red/early  and  blue/late
galaxies. In  the next   section  we   analyse  the  color  and   SSFR
distributions in our composite  sample   up  to  $z\sim  3$ to   probe
the presence   of  bimodality at  high redshifts.  The bimodality will
be used  as  a  further criterion  to  study   the evolution  of   the
luminosity functions  of the two populations.

\subsection{The bimodal color distributions}


The    inclusion  of the  deep  Hubble Deep  Fields  $H$ and  $K-$band
selected  galaxies  in  our  composite  sample   shows      that  this
bimodality continues at least up to    $z\sim     2-3$     and  mainly
separates       blue/star-forming        galaxies     from       the
red/early-type population. 

Previous  analyses  (e.g. Bell  et al.  2004) have   also  shown  that
there   is   a     weak  luminosity    dependence    of   the    color
separation   induced    by  a  color-magnitude   relation.   In  Fig.3
this  relation   is  shown   for the   two  redshift   intervals.   To
get  a first   evaluation of this relation we have  simply fitted  the
distribution shown in  Fig.3 with the  sum  of two  gaussians   having
the    mean   which is    a   linear  function  of   the blue absolute
magnitude.  Each   gaussian  has    a constant    dispersion  and each
survey has     been weighted   with its    covering sky   area.  Since
the statistics  of the  red  population  are poor  we    have  adopted
the same dependence    on  the   absolute magnitude   for both     the
blue  and    red       populations.     The     maximum     likelihood
clearly indentifies    the     locus    of     the     blue population
in  the  redshift  interval   $z=0.4-1$,  with   a  mean   color which
correlates with  the   absolute   magnitude   following   the relation
$\langle U-V    \rangle=   (-0.098\pm    0.03)    \cdot    (M_B+20)  +
(1.11\pm  0.02)$  with a  scatter  $\sigma =   0.24$.     For the  red
population   we find  $\langle U-V\rangle = (-0.098\pm    0.03)  \cdot
(M_B+20)+(1.86\pm  0.03)$ with  $\sigma   = 0.24$.

In   the       magnitude   range  present  in  each redshift
interval,     the  changes    in  the   average  colors  of  the   two
populations   are    at   most    $\Delta  (U-V)\sim      0.3$    from
$M_B=-17$    to   $M_B=             -22$.     In        the     higher
redshift   interval  $z=1.3     -3.5$      the     statistics      are
poor        and     the  significance    of    the  correlation     is
weaker.  For   the     blue  galaxies      we        derive   $\langle
U-V\rangle         =(   -0.062\pm             0.017)\cdot (M_B+20)   +
(0.73\pm     0.01)$ with   $\sigma         =0.25$         and
for      the red    population     $\langle   U            -V\rangle=(
-0.062\pm 0.017)\cdot    (M_B+20)+      (1.65\pm   0.28)$         with
$\sigma  =0.42$.    An           intrinsic   blueing   as            a
function    of  redshift       is  apparent     for  both populations,
as   expected  even     in     the       case    of     pure   passive
evolution.      At $M_B=        -20$  from    $z\sim  0.7$   to $z\sim
2.5$  it    is $\Delta   (U-V)\sim    -0.4,-0.2$      for   the   blue
and     red population, respectively.

Taking into account the  different normalizations of the  two gaussian
distributions, it is possible at  this point to find  a formal minimum
for the sum of the two distributions, which  allows to separate  in an
empirical  way the blue from the red galaxy populations. This locus is
represented by the   following   color-magnitude  relation    $\langle
U-V\rangle     =  -0.098\cdot   (M_B+20) +   1.64$   at  $z=0.4-1$ and
$\langle   U-V\rangle  =-0.062\cdot   (M_B+20)   +   1.31$  at  $z=1.3
-3.5$ (dotted  lines in  Fig. 3).  In the  low redshift  bin our $U-V$
color separation is in good  agreement with that found in  the COMBO17
survey  at $z\sim  0.5$ ( $\langle   U-V\rangle \sim   1.6$  at  $M_B=
-20$, after changing   $h$ and adopting  an average $M_B-M_V=0.5$   as
derived from the red sample). 


 The histograms  of the   color distribution  are shown  in  Fig.4
in the two redshift intervals.  The histograms have been projected  at
the average characteristic magnitude of the two Schechter  populations
in the two redshift  intervals for illustrative purpose.  The vertical
lines  represent the  average separation between the two populations evaluated
from the color bimodal distribution.

We note  in this  context that  the high  redshift interval  where the
bimodal    distribution   is     measured   is    rather    large   as
required  to  keep sufficient  statistics. Only  few red objects   are
indeed  present  in the sample    at $z>2$ and therefore we can not firmly establish
the presence of a bimodal distribution at these redshifts. However, the color
distribution  at $z>2$ (see the dotted histogram in the lower panel  of
Fig.4) is at least consistent with such a bimodality being already in place
at $z>2$.

\subsection{The bimodal $\dot m_*  /m_*$ distributions}

Considering  the  corresponding  $\dot  m_*   /m_*$  distributions,  a
bimodal   shape  appears    at    all   redshifts,    although     its
dependence   on  the   rest-frame blue   absolute  magnitude   is only
marginally significant in the  low redshift bin  and  fully consistent
with no  correlation in  the high  redshift bin  (Fig.5).  Indeed  the
same fitting procedure with two Gaussians
gives   $\langle      \log     \dot    m_*/m_*\rangle    =   (0.036\pm
0.016)     \cdot   (M_B+20)    -  (9.09\pm   0.03)$   for   the "late"
type   galaxies      and   $\langle     \log    \dot    m_*/m_*\rangle
=(0.036\pm  0.016) \cdot   (M_B+20)-(11.5\pm  0.8)$  for  the  "early"
population   in   the  redshift interval  $z=0.4-1$.   In  the maximum
likelihood  analysis  we  have  iteratively  removed  objects    above
3$\sigma$  from the  resulting Gaussian  distributions. We  also  note
that we call  "late" and  "early"  the  two main    populations of the
bimodal   distribution  without   any  specific  reference   to  their
morphological properties.

In the    higher  redshift   interval $z=1.3-3.5$   we obtain $\langle
\log   \dot  m_*/m_*  \rangle=(-0.015\pm  0.040)    \cdot     (M_B+20)
-(8.7\pm 0.1)$     for   the  "late"  population   and  $\langle  \log
\dot    m_*/m_*     \rangle     =     (-0.015\pm       0.040)    \cdot
(M_B+20) -   (11.4\pm  0.2)$    for    the  "early"     population.  A
small increase  $(\Delta     (\log        \dot m_*/m_*)\sim  0.3,0.1)$
in   the   mean    value     of     the   specific  star     formation
rate  with      increasing     redshift  is   apparent   at  $M_B=-20$
for  both        populations,    respectively.    This   increase   of
course  reflects     the   corresponding    blueing     of    the  two
populations with  increasing   redshift,  as  recently remarked   e.g.
by Rudnick et al. (2003) and Papovich et al. (2004).
 The errors  in $\log  \dot   m_*/m_*$ derived from   the spectral
fit to each galaxy SED are of the same order ($\sim 0.15$) for almost all   the
objects  in  the sample  and therefore error weighted and unweighted ML analyses
produce essentially the same result.

As in the case  of the color distribution, it is possible to  define a
minimum between the  two  gaussian  distributions  represented  by the
relation $\log \dot m_*/m_* =0.036 \cdot (M_B+20)-10.7$ at   $z=0.4-1$
and $\log \dot m_*/m_* =-0.015 \cdot (M_B+20)-10.6$ at    $z=1.3-3.5$,
weakly   dependent   on   the  absolute   magnitude.  
The histograms  of the SSFR distribution  are shown  in  Fig.6
in the two redshift intervals as in Fig.4 for illustrative purpose.

 Thus    the  much  stronger  and  more   significant  color-magnitude
 correlation   does  not  appear    to     be    mainly    due      to
 appreciable     changes   of   the specific   star  formation    rate
 with  the     rest    frame  blue     magnitude,    but more   likely
 to variations  of  some other physical         parameters.    Keeping
 in     mind   the   various  degeneracies     involved   in       the
 derivation     of   the galaxy physical    parameters    from     the
 color   fitting    to     the  synthetic  spectral     models    (see
 e.g.  Papovich,   Dickinson   \&  Ferguson  2001;   Fontana  et   al.
 2004),  it     is   interesting    to note       that   the   fitting
 analysis  suggests   dust       reddening  as   the  main responsible
 for        the  observed     color-magnitude correlation           in
 the          blue     population. Indeed  the fraction  of  the  blue
 galaxies   with  extinction  $E(B-V)>0.2$  (big  symbols  in  Fig. 3)
 increases      with      the   blue   luminosity.  In the    redshift
 interval   $z=0.4 -1$,  considering  the      blue galaxies      with
 colors    within   2$\sigma$   from  their   average  color-magnitude
 relation,    the  fraction  of dusty  galaxies  is 36\%   and    60\%
 for    galaxies  with     $M_B>-20$  and    $M_B<-20$,  respectively.
 In     the    higher   redshift  interval   $1.3<z<3.5$  the fraction
 increases  from    24\%  at    low luminosities    to 40\%  at   high
 luminosity,   but  the  statistics is admittedly poor. This is    in
 agreement        with     previous  findings        by  Shapley    et
 al.  (2001)  for   the Lyman  break galaxies at  $z\sim 3$.

For  the  red  population the change in the fraction of dusty galaxies
from faint to  bright galaxies  is only 5\%  and  it is not sufficient
to explain the  observed  color-magnitude  correlation.   However, the
fraction of red galaxies with  solar or greater than solar metallicity
is  higher  (65\%)   for high luminosity  galaxies  ($M_B<-20$)   than
for  faint   galaxies   (50\%)  in agreement  with what  is  found  in
the local early type galaxy sample (see e.g. Kuntschner et al.  2002).
This  last   effect   could    more   likely  reproduce  the  observed
correlation even at high redshift.

Although these results are dependent on several assumptions like  e.g.
the  adopted  Salpeter IMF,  they  can help  in  identifying the  main
physical  parameters  that underlie  the  observed relations.  Further
analysis in this direction is out of the scope of the present work.

Summarizing  the analysis  of this  section, we  have shown with  this
still limited sample  that the bimodality  in the distribution  of the
galaxy $U-V$  colors or   specific star  formation  rates  extends  to
high   redshifts  at  least  up   to  $z\sim   2$.  A  more     robust
statistical      significance     will  be   obtained   with    larger
multicolor optical-NIR   surveys.

Despite all  the uncertainties,  the derived  bimodal distribution  is
clearly representative of   the    presence     of     two    distinct
populations   and    it   can   be    used   to    make   an empirical
separation  of  the  galaxy  populations  at  different   redshifts to
analyze their luminosity   function evolutions. This  will be done  in
the next section.

\section{The evolution  of the luminosity functions  of field galaxies}

\subsection{The statistical analysis}

The luminosity function  is  computed extending the procedure  used in
paper  I.  First  of all   we have    applied   to     our  composite
sample an extended   version  of the     $1/V_{max}$  algorithm   
as  defined  in Avni and  Bahcall (1980) so that  several samples  can
be combined  in  one  calculation. Our     combination    of     data
from  separate     fields,    with different   magnitude   limits, has
been treated  computing,     for every     single       object,      a
set of    effective  volumes,  $V_{max}(j)$, for each  $j$-th    field
under analysis.   For   a  given  redshift interval   $(z_1,z_2)$,
these   volumes  are  enclosed  between  $z_1$   and  $z_{up}(j)$, the
latter being  defined as  the  minimum  between $z_2$ and the  maximum
redshift at  which the  object  could have been   observed within  the
magnitude  limit  of  the  $j$-th  field.  The   galaxy  number
density $\phi(M,z)$ in each  $(\Delta  z,\Delta M)$  bin can then   be
obtained as:

\begin{equation}
\phi(M,z)=
\frac{1}{\Delta M}\sum_{i=1}^{n}\left[
\sum_j \omega(j)\int_{z_{1}}^{z_{up}(i,j)}
\frac{dV}{dz}dz
\right]^{-1}
\end{equation}

where   $\omega(j)$   is   the   area   in    units   of    steradians
corresponding to the  field $j$, $n$   is the  number   of objects  in
the  chosen bin  and $dV/dz$  is the comoving  volume element per
steradian.  

The  Poisson error  in  each LF  magnitude  bin was  computed adopting
the   recipe  by   Gehrels  (1986), valid   also  for  small  numbers.
The   uncertainties   in  the  LF  due  to     the     photometric $z$
estimates  were computed  as in  paper I,  assuming    the     typical
r.m.s.     found comparing   photometric     and         spectroscopic
$z$.  A       set   of    50   catalogs   of  random   photometric $z$
was produced using the  above   r.m.s.  uncertainties to       compute
a   set  of  50   LFs.   The   derived        fluctuations  in    each
magnitude  bin resulted smaller   than  Poissonian  errors and    have
been   added   in   quadrature    although  the   errors  are  not
independent  between  different  bins  and  could  be   overestimated.
Finally,  in  order to  take  into account,  at  least partially,  the
uncertainties  due to the  field-to-field variance,  we have used  the
maximum between the Poisson error and the density  variance  among the
fields   in  each  bin   of  the  luminosity  function.  The resulting
field-to-field variance is the main source of error in the LF bin.

The  $1/V_{max}$ estimator for the LF can be affected in principle  by
small scale galaxy clustering although in our case the large  redshift
bins adopted would reduce the effect of any  density fluctuation.  For
this reason  a parametric maximum likelihood estimator is also adopted
which is known  to be less  biased respect to   small scale clustering
(see Heyl et  al. 1997).

The parametric analysis   of  the   galaxy  LF has been  obtained from
the   maximum     likelihood    analysis    assuming   a     Schechter
parametric   form   $\phi$    for   the    LF.  The method used in
Poli et al. (2003) and in the present paper represents an extension of
the standard   Sandage,   Tammann    \&  Yahil    1979  method   where
several samples can be  jointly analyzed and  where the LF  is allowed
to  vary with  redshift. In the  case of  our  composite survey,  that
is     based  on    $j$    magnitude-limited   fields,   the   maximum
likelihood $\Lambda$ has been  derived from  the analysis  of Marshall
et  al.  (1983)  after  proper  normalization  of  their   probability
function: 

\begin{equation}    
\Lambda   =
\prod_{i=1}^{N}\frac                            {\phi(M_i,z_i)dzdM}{\sum_j
\omega(j)\int_{z_1^j}^{z_2^j}\frac{dV}{dz}dz       \int_{-\infty}^       {
M_{lim}^{j}(z) }\phi(M,z)dM} 
\end{equation}

where $z_1$,$z_2$ are the minimum and maximum redshifts available for each
sample,
$N$  is the    total number     of    objects    considered   in
the  composite     sample,  $M_{lim}^{j}(z)$  is    the   absolute
magnitude value    corresponding,  at   the   given    redshift   $z$,
to     the  magnitude limit    of   the  $j$-th survey.   The value of
$\phi^{*}$  in Table  1  is  then     obtained  simply        imposing
on       the best      fit       LF     a normalization such that  the
total  number of   galaxies  of  the combined  sample  is  reproduced.
  A formal derivation of this equation can also  be found in  Heyl
et al.    (1997) where   it is  also shown  that the present  equation
reduces to  the   standard   Sandage     et    al. formulation  in the
case  of  a   single survey   and   of  a LF   which  is   constant in
redshift.

At variance with paper I we allow the Schechter parameters $\phi^*$ and $M^*$
to vary with redshift:
$$
\phi(M,z)=0.4 \cdot \ln(10) \phi^*(z) [10^{0.4(M^*(z)-M)}]^{1+\alpha}\exp[-10^{0.4(M^*(z)-M)}]
$$
where  $\phi^*=\phi^*_0\cdot  (1+z)^\gamma$  and  $M^*(z)=M^*_0-\delta
\cdot  \log(1+z)$  equivalent  in  luminosity  to   $L^*(z)=L^*_0\cdot
(1+z)^{0.4\delta}$. The slope $\alpha$ is kept constant with redshift
since it  is well  constrained only  by the  low-intermediate redshift
data ($z<1$).

We used   the MINUIT  package  of  the CERN   library (James \&  Roos
1995) to  minimize $-2ln\Lambda$.   Since both  the composite  sample
and the parametric  analysis have   been improved  respect to   paper
I, we show  in Fig.7 and  in the first   row  of Table  1 the results
about  the evolution  of the  luminosity function   of the    overall
sample.   The  errors  in  Table  1  correspond  to  errors  for each
interesting  parameter,  independently  of   the   other  parameters,
obtained  with  a standard threshold  of  $2\Delta  ln\Lambda=1$.   A
detail  analysis  of  the 4-dimensional space is out of the scope  of
the present paper.  However, if  we  consider  e.g.  the  joint  68\%
confidence  region   for  the  $M^*,\alpha$ or  $\gamma,\delta$ pairs
of   interesting  parameters  we   obtain,   after projection on  the
parameter axes, individual errors that are about 1.6 times larger. 

The inclusion  of  the  deep NIR   selected  HDF-N sample  contributes
to  a   better definition   of  the  faint  LF  slope at  intermediate
redshifts  ($z=1.3-2.5$).  As  in  paper  I,  the  local  LFs from the
Sloan (Blanton  et al. 2001)  and  2dF (Norberg et al.  2002)  surveys
have  been  included for   comparison. The   difference   between  the
two   local  LFs  show   the  level   of uncertainty present   in  the
estimate  of    the  LF   even   at  about  the  same  local  redshift
interval.

The evolution of the  LF  shown in Fig.7  by the  $1/V_{max}$ analysis
shows little density   evolution at the   faint end  with  respect  to
the local  values, while   at the  bright end  a brightening  by about
1 magnitude is apparent.

If we adopt a Schechter shape of the LF at all redshifts, the  maximum
likelihood analysis shows an  average  slope, ($\alpha \simeq  -1.3$),
somewhat  flatter     than  that   found   in   Paper  I.    Moreover,
appreciable   density  and  luminosity  evolution   with  redshift are
needed to fit the data,  as  derived from the parameters in Table   1.
 We note in this respect that the magnitude limits of the  samples
cut  the  LF  at  progressively  higher  luminosities  with increasing
redshifts.  For  this reason  we  prefer to  keep  the $\alpha$  slope
constant with redshift unless a  bad agreement with the 1/Vmax  points
is found.   The mixed   luminosity-density shift   of the   Schechter
shape  provides   a   luminosity        function    which   keeps   an
almost constant  volume density   with   redshift      at  the   faint
end, while     at     the   bright      end         it        provides
an appreciable     luminosity evolution,      in    agreement     with
the qualitative    results  of the  $1/V_{max}$ analysis.

Fig.7  also shows  the good  agreement, better  than the 2$\sigma$
level, between the   parametric and  the $1/V_{max}$ LFs. This  implies
in  particular that  the  assumed  parametric  evolutionary  laws  are
an acceptable  representation  of   the  binned   luminosity  function
in   the  overall  redshift  interval.  Moreover,  the  good agreement
implies that any  small scale clustering  does not affect  appreciably
the $1/V_{max}$ estimate of the LF due to the large redshift intervals
adopted. 

This kind  of shape  and evolution  of the  luminosity function is the
result  of   the  different   evolutions  of   distinct   galaxy
populations.  For  this  reason  we  perform  in  the  next  section a
separation  of  the  galaxy  sample  in  the  blue/late  and red/early
populations to probe the  different evolutionary behaviour of  the two
luminosity functions.

\subsection{Passive  evolutionary tracks  as a  criterion to  separate
red/early and blue/late galaxies as a function of z}

The luminosity functions  of the blue  and red populations  defined on
the basis  of the  "S0 color track"  are shown  in Fig.8 in   different
redshift  intervals.   The parametric  Schechter  evolution has   been
derived adopting  a mixed  luminosity  and  density evolution  with  a
constant  slope.  The   model luminosity  function   is  shown at  the
average $z$ of the bin for illustrative   purpose. 

The  results   of  the   Maximum   Likelihood   fits  are   shown   in
Table 1.  Assuming the  Schechter shape at  all redshifts we  found that
the  average  slope  $\alpha$   of    the    two    populations     is
clearly  different,  ranging  from   $\alpha \sim -1.3,-1.4$ for   the
blue/late galaxies  to $\alpha   \sim  -0.5,-0.6$   for the  red/early
population.  This result is already well known from previous works  on
the   local  and  $z<1$    samples   (Lilly   et   al.   1995, Blanton
et al. 2001,   Bell  et    al.  2004). The shallower  slope   for  the
red/early  sample  implies that  the ratio  of the  blue over  the red
population increases strongly to fainter magnitudes. The slope of  the
red  LF in particular is in good agreement with the value of  $-0.6\pm
0.1$ derived  in  the  $z=0-1$ interval   by  Bell  et   al.    (2004)
in their analysis  of  the  wider   but shallower COMBO17 survey.  The
$\delta$  and  $\gamma$  parameters  of  the  luminosity  and  density
evolution  are significantly  different from  0 suggesting  that both
evolutions are needed to provide a satisfactory fit.
Both populations  show comparable luminosity evolution, but   the  red
population is  characterized by  a stronger  density evolution  with a
decline by at least   a factor  $\sim 5$  from $z\sim  0.4$ to  $z\sim
3$.

A comparison with the shape and evolution of the total LF computed  in
the previous section (see Table 1) shows that both the faint end slope
and  the  amount  of  density  evolution  of  the  total LF are mainly
driven by that of the blue population.

To  explore  how  much  the  evolution  of  the  red  galaxies  can be
influenced by the presence in this population of dusty starburst galaxies, especially  at high
$z$, we have  separated the population  according to the  evolutionary
"S0 track" of  the specific  star formation  rate $\dot  m_*/m_*$.

The resulting luminosity  functions are similar to the  previous  ones
with  only   mild  differences  (Fig.9). From the last column of Table
1, where the number  of objects  in each  sample is given, it   can be
noted that  about  35\% of the  red sample with  colors  redder   than
the  "S0  color  track"      shows   a  spectral  energy  distribution
indicative  of   a  dusty  star   forming population.  This   fraction
is   similar   to   what   is  found    in spectroscopic  surveys  and
does    not    alter    the    qualitative  behaviour  about   the LF
evolutions.

It  is   clear  that   the  presence  of  a  mixed  luminosity-density
evolution  in  both  populations  implies,  in  particular   for   the
red/early  population, a   star formation  history more  complex  than
provided   by  the  simple  PLE   model  adopted  here to separate the
two  populations, where       the burst     of   star   formation   is
located    at   high    redshifts  ($z>3$).  This  mixed evolution  is
directly seen in the  $1/V_{max}$  representation of the LFs in  Fig.8
and Fig.9, which shows that  luminosity    evolution     is   apparent
mainly      at  the bright   end     of   the   LFs,   while   at  the
faint         end      the   LFs       appear        affected       by
density      evolution.      This    behaviour       will           be
discussed    more quantitatively  in section 6.

\subsection{The empirical bimodal color  and $\dot  m_*/m_*$  distributions  as
a  criterion  to   separate red/early and blue/late galaxies as  a
function of z}

In the previous section we have shown that the evolutionary  behaviour
of  the red/early  population does  not follow  what expected  by  PLE
models,    assuming  a    passive  evolutionary   scenario  for    the
color or $\dot   m_*/m_*$ separation. A  selection criterion which  is
independent  of  the assumed  evolutionary  scenario is  based  on the
empirical analysis  of the   color or  $\dot  m_*/m_*$   distributions
as observed  in  our  sample and  already found  at lower redshifts by
Bell  et   al.  (2004) and  Baldry et  al. (2004).  In the   previous
section  we  have shown that these  distributions are   bimodal up  to
$z\lesssim  3$ and  so identify  two distinct  populations. The  peaks
of  the bimodal   color distributions derived  in  the  low   and high
redshift   intervals   are  dependent     on  both      redshift   and
luminosity,   while    the  $\dot   m_*/m_*$  distributions  are  less
dependent  on luminosity.

Using  the locus   of the  minimum  in  the color  and $\dot  m_*/m_*$
distributions in   the $0.4<z<1$  and $1.3<z<3.5$  intervals, we  have
derived the  evolution  of the luminosity  function  of red/early  and
blue/late galaxies shown    in Fig.10 and   Fig. 11, whose   parameters
are shown   in  Table   2.  In   this case,   in  each   of   the  two
redshift  intervals, the  color  separation  is   a  function  of  the
rest-frame  blue absolute  magnitude.  At  $M_B=-20$ the   color that
divides  the  two classes   becomes redder   by  0.3  mag  from $z\sim
2.5$  to $z\sim 0.7$, similar although a bit smaller  than predicted by
the PLE  color track used in the previous section.

For the red population  we have compared our  LF with that of  Bell et
al.  (2004)  finding a  general  good agreement  at  relatively bright
magnitudes and  low redshifts.  In particular,  the slope   of the red
population is $\alpha = -0.76\pm  0.10$ and  even in  this case it  is
in  agreement with  that   found  in the COMBO17 survey. 

We  do not    consider  the effects    of  uncertainties in  the
location of  the  minima of the  bimodal distributions, but  we do not
find a  change in the qualitative evolutionary   behavior of the   two
populations  comparing   the  LF   results  obtained   using  the   S0
evolutionary  track and the minima in the bimodal distributions.   The
main   quantitative   difference   consists  in  a   stronger  density
evolution  (larger $\gamma$)  for  the  red/early population  in   the
bimodal  case  (since we are  excluding some objects at high redshifts
with respect to the "S0 cut"). 

Indeed we  can compare   the luminosity function     of     the  early
population selected  to  have $\log  \dot  m_*/m_*\lesssim    -10.5,  
-10.6$         nearby          the        characteristic      absolute
magnitude     $M^*=-21,-21.5$     (depending        on redshift).   At
these    magnitudes  we    are     on    the flat     part  of     the
luminosity    function  and   thus       we     are     less sensitive
to   changes       in  volume  density     due     to   any luminosity
evolution.  In   Fig.11  we can    see   that     the   LF  of     the
"early"   sample     declines  by  about    one    order of  magnitude
from   $z=0.5$     to   $z=3$.   This  density   evolution is somewhat
smaller     when      considering     the   color-selected  red sample
because   of    the  presence   of   some  dusty   starburst  galaxies
at high redshifts.  Indeed $\sim 40$\%  of the red   sample shows SEDs
indicative of a dusty star-forming population.

\section{Number and luminosity density evolutions of the  red/early
and blue/late galaxies}

\subsection{ Observed behaviour}

The values of the parameters of the mixed evolution  we have   derived
for    both   populations   are    the   result   of    the    assumed
parameterization of  the LF   with a Schechter function.   To  clarify
in  a  less  model-dependent  way  the   cosmological   evolution   of
galaxies  at different luminosities we  evaluate  directly  the volume
density  of galaxies  as a  function   of redshift  in  given bins  of
the rest-frame blue  magnitude. We can   also evaluate  the associated
redshift  evolution   of   the luminosity   density  observed   in the
rest-frame  B band by  the  same galaxies.

We  show  the results    both  assuming the "S0   evolutionary  track"
and  the  empirical  separation derived from the bimodal distribution 
to  separate  the   two populations.

Considering  the    former   case    we note     in  Fig.    12   that
the cosmological  evolution   of    the   red/early  galaxies  appears
to  depend  on   their  luminosity.   Indeed   the  faint    galaxies,
typically   in  the   range    $-21.5<M_B<-20.6$  (where    $M_B\simeq
-20.6$  is  at  the faintest limit  of the population   at $z\sim 3$),
decrease  by  a  factor  $\sim  2-3$   from  $z\sim 0.5$    to  $z\sim
2.5$.    However   it  is  interesting   to  note  that the   brighter
galaxies  in the   range   $-23<M_B<  -22$ are  consistent with  being
constant in the same redshift interval.  

To    allow   a   first   simple quantitative      comparison  with  a
PLE   scenario,    we       also      show   in      Fig.12        the
expected   evolution  for   galaxies   with  $  -23<M_B<-22$  obtained
adopting      the   local         LF         of  red          galaxies
and evolving   it in luminosity   adopting    the "S0     evolutionary
track"  shown      in      Fig.1. The  local value    derived     from
the  Sloan   survey   is  included  as   derived from    Bell   et al.
(2004).    Although  the   COMBO17     sample       has   been derived
from       a    slightly   different          color    threshold,  the
intermediate  redshift   values     are      in  good  agreement  with
our    data. 

The PLE behaviour is    similar to the observed evolution  for  $z<1$,
but it  overestimates   by       a  factor   $\sim   5$    the    data
at high  redshifts.  This  overestimate reduces  to a factor $\lesssim 3$ if
compared with the  number density   derived from  the best  fit  LFs.
This  would  imply   that      dynamical  events    are important  for
the brightest red/early   population at $z>1$.

The B-band luminosity density derived   from  the same data is   shown
in  Fig.13   and   has   the    same    behaviour   of    the   number
density    evolution.    The  data   (open   circles)  represent   the
contribution    from red/early  galaxies   with  $M_B<   -20.6$  which
are  at   the faintest limit    at  $z\sim  3$.    In  this    way
we   can  explore  the evolutionary behaviour over the entire redshift
range of galaxies within the  same luminosity range. At $z<1$  we have
also  considered  (open  triangles) a   lower   limit   $M_B<-19.5$ to
compare the  luminosity  density   at intermediate  redshift with that
of  Bell    et al.    (2004) finding  a broad  consistency within  the
rather large  errors. Our  values however  are about  a factor
$\sim 2$ higher since the "S0 track" used to separate the two populations
tends to include in the red population a higher number of galaxies with increasing redshifts
respect to the bimodal class separation used by Bell et al. (2004).
The    same
PLE model   is also   shown  in  Fig.13 and  the same   discrepancy is
apparent at $z>1$.

Since  the   evolution  is   more  complex   than  expected   from   a
PLE scenario, it is interesting to use the empirical criterion for the
class    separation     based    on    the   star-formation    bimodal
distribution. The   volume   densities   and  the  B-band luminosity
densities of galaxies with   $M_B<-20.6$ as   a function   of redshift
are shown   in Fig.14.  In  this  case   both  the   early  and   late
populations   are shown in each panel.

The    cosmological    evolutions    of    the   bright  "late"    and
"early" populations appear  to  have a  bifurcation at  $z>1$.  Indeed
while      star-forming      galaxies    slightly   increase  or  keep
constant  their  number/luminosity    density,     "early"    galaxies
decrease  in   their number  density by  about one  order of magnitude
and  in  luminosity  density by  a  significant  factor   $\sim   5-6$
(although with   large  statistical     uncertainties)  from    $z\sim
0.4$   to $z\sim    2.5-3$.  We   have   also plotted  the  luminosity
density    of  both  populations  derived   from   the   best  fit LFs
extrapolated down to  the minimum  luminosity observed in  our low $z$
LF, $M_B=-15.8$.  The   evolutionary   trend   is   similar  to   that
observed  although  shifted    in   normalization   by   a      factor
$\lesssim   2$.   It  is  important   to  note  that   the  parametric
representation  used   in the  present analysis    is valid   only  in
the adopted   redshift interval $0.4<z< 3.5$.  Extending the  analysis
outside  this redshift interval could require more complex  parametric
evolutions.  In     any       case,  while      at       $z<1$    both
populations              give            a  comparable    contribution
to   the      total  B-band     luminosity density,     at       $z>1$
the   contribution     by    the       red population     is      much
lower.   This   results     in       a  global  luminosity     density
of   galaxies  with    $M_B<-20.6$ almost constant     with   redshift
in the   same $z$ interval.

Since  we    have  seen   in  paper    I  that   there    is  a strong
correlation between  the observed   rest-frame  blue   luminosity  and
the   star-formation   rate  for     the      blue/late    population,
this   implies    that   the star-formation activity in   star-forming
galaxies slightly increases or keeps constant up to  $z=3$.   On   the
contrary,    the number   of  galaxies     with low   specific   star
-formation        rate  $\log      \dot     m_*/m_*<    -10.5$   which
are     representative  of       an      old       stellar  population
with  high     formation redshift $z>3$, decreases quickly.

This behaviour is in qualitative  agreement with what expected from  a
hierarchical  clustering  scenario where  galaxies  with low  specific
star-formation are located in the rare peaks of high dark matter density
at  high redshifts. However, a detailed  quantitative comparison  with
specific  CDM  models  is  required   to  understand  which   physical
mechanisms govern the early evolution of galaxies. A first  attempt is
done  in the next section.

\subsection{Predictions by CDM models}

To  understand how  the above  results fit  into current  hierarchical
models  of  galaxy formation  we used our  semi-analytic  model (SAM).
Starting  from  primordial  density  perturbations in  a $\Lambda$-CDM
Universe, the model describes  the collapse and the  merging histories
of the DM haloes which originate from such perturbations.
In  previous
papers (see, e.g., Menci et al. 2002) we have described   our model in
detail; it includes  the dynamical
friction   and the  binary  aggregations,  determining  the  fate   of
clumps  included into  larger  DM  haloes   along  the  hierarchy   of
mergers.
The baryonic  processes  (gas   cooling, star   formation,
Supernovae feedback)  are linked  to  the dynamical  history  of  the
galaxies   through  the  simple  recipes  adopted  by  current    SAMs
(Kauffmann et  al. 1993;  Cole  et  al. 1994;  Somerville  \&  Primack
1999; Poli   et al.   1999; Wu,   Fabian \&  Nulsen 2000;  Cole et al.
2000). 

In this paper  we are interested  in the whole  distribution of colors
for any given galactic mass  and redshift. In turn, such  distribution
depends on the different galaxy  merging histories leading to a  given
galactic  mass.  Thus,  we  updated our  model  to  describe  the {\it
scatter} in the galactic properties (star formation rate, colors)  due
to different  merging histories,  which could  not be treated with our
earlier model (Menci et al. 2004). To this aim, we built a Monte Carlo
version  of our  SAM, where   different histories   are generated   by
extracting  the merging probability according to  the prescription  of
the extended Press-Schechter (EPS) theory; the corresponding galactic mass function is computed 
from such numerical realization (at variance with the method used in Menci 
et al. 2002; 2004).

The  properties of  the gas  and stars  contained in  the galactic  DM
clumps are computed  as in  Menci et al. (2004). The  only differences
are that now i)  stars are  allowed to   form with  rate $\dot   m_* =
m_c/ t_{*}$,   with  the   star  formation   timescale  $t_*$  assumed
to   be  proportional to the disk   dynamical   time  $t_*=q\,r_d/v_d$
through the  proportionality constant  $q$, which  is left  as a  free
parameter; ii) the  mass $\Delta  m_h$  returned from the  cool to the
hot  phase  is  estimated  as  $\Delta   m_h=E_{SN}/v^2_{esc}$   where
$E_{SN}=10^{51}\,{\rm    ergs}\,\epsilon_0\,\eta_0\,\Delta    m_*$  is
computed  from  the  mass  $\Delta  m_*$  of  stars  formed   assuming
$\eta_0\approx (3-5)\cdot 10^{-3}$  as the number  of SNae per  unit solar
mass (depending  on the   assumed IMF),  and $\epsilon_0=0.01-0.5$  as
the efficiency  of the  energy transfer  to the  interstellar   medium
whose  value  is  highly  uncertain (see,  e.g,  Kauffmann,  White  \&
Guiderdoni 1993; Somerville \& Primack 1999) and it is considered as a
free  parameter.  The  model   free  parameters   $q$ and $\epsilon_0$
are chosen as to   match the local B-band galaxy  and the Tully-Fisher
relation.  We   adopt  a    Salpeter  IMF   with  a   standard   value
$\epsilon_0=0.1$    

At each merging event, the masses of the different baryonic phases are
replenished by those  in the merging  partner; the further  increments
$\Delta m_c$, $\Delta m_*$, $\Delta m_h$ from cooling, star  formation
and  feedback  are  recomputed on  iterating  the  procedure described
above.  In  our  MonteCarlo code  we  adopt  the  description of
starbursts triggered   by  major/minor  merging  and  galactic  fly-by
following the formulation in Menci et al. (2004).

At  higher   $z$,  the  model results  match  the  observed  K-band LF
(Pozzetti et al.  2002) up to $z=1.5$   and the UV  LFs  measured  for
Lyman-break galaxies  at $z=3-4$.  The  global  star  formation  history
predicted by the  model is consistent  with   that  observed   up   to
redshift  $z\approx  4-5$ (Fontana  et al. 2003a;  Giavalisco  et  al.
2004),  and  the  predicted stellar mass  functions fit  the  observed
ones (Cole   et al.  2001; Fontana et al. 2004) up to $z=1.5$. 

Given the above agreement  concerning the {\it average}  properties of
the  galaxy population  (i.e., the  properties averaged  over all  the
merging histories) the  comparison of the  SAM results with  the LF of
populations with different  $\dot m_*/m_*$ constitutes  a higher-order
probe of  the formation  history of  galaxies which  allows us to test
also the properties of populations with particular merging histories. 

The  predictions  by our  CDM  model concerning number and  luminosity
densities are  shown  in Fig.   14  for both populations   (late   and
early)    adopting      the     same      empirical    separation
derived  from    the   observed    bimodal   distribution. Indeed  the
model shows a  clear   bimodal distribution pointing toward   the same
color dichotomy  found in  the data. A detailed analysis of the  model
bimodal distribution will be presented in a subsequent paper.  It  is
to  note that  since the    red/early  luminosity    functions at  all
redshifts are rather flat around $M_B\sim    -20.5$,  the   comparison
with     the model  predictions depends only weakly on the magnitude cut.

We have to keep  in mind that the  threshold  in the $\dot    m_*/m_*$
adopted  to  separate    the two   populations  could  be subject   to
systematic  underestimates   especially concerning  the   stellar mass
derived   from   the   observed    SEDs,  as  discussed  in  sect.  3;
  in particular, the "maximal age approach" yields stellar  masses
a  factor  2 larger  (Fontana  et  al. 2004);  thus we conservatively
adopt in  the  predicted distribution of  $log(\dot m_*/m_*)$ a  spread
ranging from  the default   threshold  (-10.5,   corresponding to  the
best  fit observed stellar mass estimate) up to an offset of 0.3.  The
assumed  uncertainty  in the  threshold  propagates   to  the    model
predictions     concerning    the   number/luminosity   densities,  as
represented  by the  shaded   regions in  fig.   14.  In   principle,
an  additional  source      of    uncertainty     is  represented   by
modelling    the dust     extinction   curve,    which  affects    the
derived     SFR.  However,   for     galaxies    with   $\dot m_*/m_*$
around the  threshold value,   the low  amount  of    dust  makes this
uncertainty much   lower than that related to the stellar mass.

Figure 14  shows a  broad agreement  between the  observed and     the
predicted   evolutions of   the early  and late   type galaxies   with
$M_B<  -20.6$. In  particular, a   decrease in  the  number/luminosity
density  of red/early  galaxies by   a   factor   $3-4$   respect   to
the  local   value  is predicted  at $2<z<3$.   Thus,   finding    red
objects       at   high  $z$     is      not    inconsistent  $-$   by
itself $-$ with    hierarchical  clustering      scenarios.       This
is  because         the    clumps   which      end        up        in
massive, bright      objects        are   predicted       to      have
older stellar  populations       (and hence  redder      colors)    on
average  since   they       formed       in   rare,     high   density
(biased) regions     at high       redshifts. The   higher   densities
present  in   these    regions    have    provided  enhanced      star
formation, rapidly   exhausting  the cold   gas reservoir.

Thus, at  the   level   of details of     the   present    SAMs     it
is  remarkable   that    a     clear  distinction     between      the
evolutionary     trends     of      the     two   populations       is
predicted            by  this    class       of  models.   A  detailed
study of  the   link    between the    different merging     histories
leading      to        the     two    different  populations  will  be
presented in a future paper.

A  similar    behaviour  is   also  expected   in   the   hydrodynamic
simulations  by Nagamine  et  al.  (2004) where an  appreciable number
of massive (and hence mostly red) galaxies is predicted at $z\sim 3$.

In    this   comparison     we    should  however   keep   in     mind
that several  uncertainties    affect   also  the      data    points.
Field to  field variance   due to    clustering      could      affect
in   particular       the         number              of         early
type  galaxies   observed   at        high      redshifts           in
small fields.  Moreover, the      average  fraction     of       early
type  galaxies in     the red    population  ($\sim     60$\%;    see
last  column  of    Table  2)       is   statistical       in   nature
and   based    on       the    fitting      of         the         SED
distribution    with      the      spectral    synthesis       models.
However,      unless  this  fraction  results       much     different
after       detailed spectroscopic  analysis           of          the
red      sample,  the conclusion  is that  the present  CDM models are
able to predict  the presence of a  number of early  type galaxies  up
to $z=3.5$ which  is comparable to that observed.

\section{Summary}

We  have  used a  composite  sample of  $I,H,K$  selected galaxies  to
derive the evolution of   the  B-band luminosity  function   of galaxies
separated  by their  rest-frame $U-V$   color or  specific  (i.e.  per
unit mass) star-formation rate up to $z\sim 3$.

The observed color  or SSFR distributions  appear bimodal up  to $z\sim
2.5-3$, extending to high  redshift the same behaviour  recently found
at low  and  intermediate  redshifts.  In  particular,  the  bimodality
in  the SSFR  distribution depends  weakly on  the luminosity  of the
sources in the luminosity  interval probed by  the present sample  and
evolves little from $z\sim 0.5$ to $z\sim 3$.

We adopted  two different  criteria  to  classify the  galaxies and to 
study the evolution of the luminosity function  of  the  red/early and
blue/late populations.

Focusing on  the evolution  of  the  early galaxies,  we first adopted
the null hypothesis that bright red/early galaxies are mainly affected
by substantial passive luminosity evolution. For this reason  we  have
adopted    the   color/SSFR     passive   evolutionary    tracks of  a
galaxy whose local color is similar to that of a  typical S0   galaxy,
to  separate  the    two populations   as   a  function  of  redshift.
The     resulting  qualitative   behaviour      of     the  luminosity
function      does   not     depend  significantly  on  the  parameter
used  for the  selection, $U-V$  color  or  SSFR, implying   that the
contamination  of dusty  star-forming  galaxies  in the   red sample,
although  substantial  in  number,  does  not modify  appreciably  the
global evolutionary properties  of the population.  

Adopting  a   Schechter  shape,   the   LF  slope  of  the   blue/late
population  is $\sim -1.4$  up   to   $z\sim  2.5$,  while  that    of
the      red/early  population     is    flatter     ($\sim     -0.5$)
extending    to  higher  redshifts the    similar   difference   found
at  $z<1$.  The evolution of   the luminosity    function   of    blue
and  red  galaxies   is  described    in    parametric    form      by
the presence      of  both  luminosity    and    density    evolution.
The   evolution    of  the     blue      population     is      mainly
driven    by  luminosity evolution,   while     appreciable    density
evolution characterizes the  red/early population. In    particular, a
decline by   a  factor of the  order  of  5 results  from   $z\sim
0.4$     to   $z\sim  3$.   This  implies   that      the    red/early
population   as    a   whole  can  not be     simply   described    by
simple  pure  luminosity   evolution models, but major dynamical events
could act  at high redshifts.

For this reason we also adopted an empirical criterion to separate the
two galaxy   populations, based  on  the  presence of   a  minimum  in
the  bimodal  observed   color/SSFR  distribution   in  each  redshift
interval. The evolutionary difference  between the two populations  is
even  more marked in this  case, with a  density decline of  the early
population by  about  one  order  of  magnitude  from  $z\sim  0.4$ to
$z\sim 3$  at $M_B\sim -21,-21.5$.

The  amount  of  mixed  luminosity-density  evolution  depends  on the
assumed parameterization of  the LF. We  have also evaluated  directly
the  volume  and  luminosity  density of  galaxies  as  a  function of
redshift in given  intervals of absolute  magnitudes. We have  adopted
both  selection   criteria to  separate  the  two galaxy  populations.
Assuming  the  "S0  passive  evolutionary  cut"  we  have  found  that
while   the  faint   red/early  population  slightly  decreases   with
redshift, the brightest  fraction   with   $M_B<-22$   is   consistent
with   being constant     with redshift  up   to  $z\sim 3$.  A simple
PLE model where galaxies are formed  at high redshifts   overestimates
by a   factor of the order of 5 the bright fraction at $z\sim 3$.

Adopting the  empirical criterion  we have  found that  the volume and
luminosity  densities  of   the early   and   late  populations having
$M_B< -20.6$,  which is   the faint  limit  in  the highest   redshift
bin,   appear   to  have   a   bifurcation  at  $z>1$. In  particular,
while  star-forming       galaxies     slightly   increase  or    keep
constant their luminosity    density,    "early"   galaxies   decrease
in their  luminosity  density   by    a   factor    $\sim   5-6$  from
$z\sim 0.4$  to $z\sim 2.5-3$.

A comparison  with  our   latest version  of hierarchical  CDM  models
shows  that  a clear  distinction between  the different  evolutionary
histories  of  the   early  and  late   type  galaxies  is  predicted,
in    good    agreement    with    the    observations   where     the
contribution in the luminosity density of the early population is  $10
-20$ times below  that  of the   late population  at   $z\sim  3$.  It
is   clear      however       that      a        more         detailed
comparison  requires      larger   surveys      to   reduce        the
effects      of  the      field-to-field   variance   and     at   the
same  time      a better     description    in   the   CDM      models
of     the  physical  mechanisms      governing        the   different
cosmic histories   of the two populations.

\acknowledgments {We thank the  anonymous referee for her/his  helpful
comments. We  also thank A.   Cimatti for  the use  of the K20  survey
data.}

\clearpage

\begin{table} 
\begin{center}
\caption{Luminosity function parameters}
\begin{tabular}{ccccccc}
  \hline
    --- & $\phi^*_0$  & $M^*_0$& $\alpha$ & $\delta$ & $\gamma$ & N \\
  \hline

  Total             &  0.0056   & $-20.93\pm0.23$  & $-1.29\pm0.04$  & $1.48\pm0.52$   &$-0.65 \pm 0.19  $& 1434 \\   
  $<U-V(S0)$        &  0.0042   & $-20.31\pm0.26$  & $-1.38\pm0.05$  & $2.35\pm0.58$   &  $-0.52\pm 0.23$  & 1142 \\
  $>U-V(S0)$        &  0.0106   & $-20.23\pm0.28$  & $-0.46\pm0.10$  & $2.72\pm 0.66$  & $-2.23 \pm0.28$ & 292  \\
  $>\dot{m}/m(S0)$  &  0.0050   & $-20.55\pm0.25$  & $-1.33\pm0.05$  & $1.96\pm0.56$   &  $-0.60\pm 0.22$  & 1242  \\
  $<\dot{m}/m(S0)$  &  0.0044   & $-20.66\pm0.36$  & $-0.63\pm0.11$  & $2.08\pm0.82$   & $-1.80\pm0.34$  & 192  \\

  \hline
 \end{tabular}
\end{center}
\end{table}

\begin{table} 
\begin{center}
\caption{Luminosity function parameters}
\begin{tabular}{ccccccc}
  \hline
    --- & $\phi^*_0$  & $M^*_0$& $\alpha$ & $\delta$ & $\gamma$ & N \\
  \hline
  
  $<U-V(bimodal)$        &  0.0040  & $-20.47\pm0.26$  & $-1.36\pm0.05$  & $2.30\pm0.57$   &  $-0.45\pm 0.22$  & 1173 \\
  $>U-V(bimodal)$        &  0.0110   & $-20.49\pm0.38$  & $-0.76\pm0.10$  & $2.70\pm 1.00$  & $-3.10 \pm0.38$ & 261  \\
  $>\dot{m}/m(bimodal)$  &  0.0050   & $-20.51\pm0.24$  & $-1.32\pm0.05$  & $2.15\pm 0.53$ &$-0.56\pm 0.20$  & 1282 \\
  $<\dot{m}/m(bimodal)$  &  0.0110   & $-20.23\pm0.41$  & $-0.53\pm0.13$  & $3.49\pm 1.01$  &$-3.65\pm 0.47$  & 152 \\

  \hline
 \end{tabular}
\end{center}
\end{table}


\clearpage

\begin{figure}
\plotone{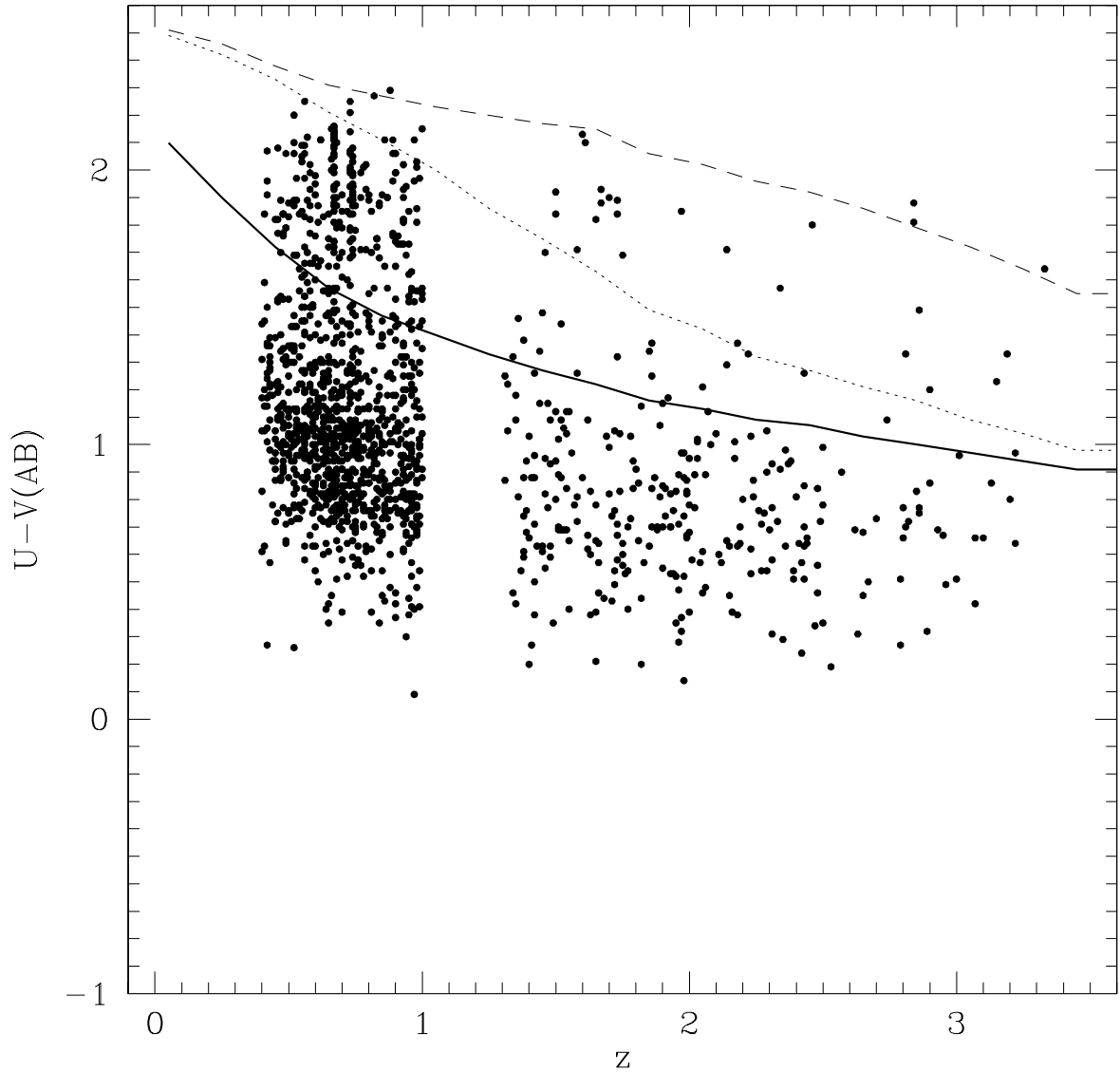}
\caption{Rest-frame $U-V$ as a function of redshift. The continuous curve represents the color evolution due to the
passive evolution of a galaxy formed at $z=10$ and with a star-formation time-scale $\tau = 3$ Gyr. The Bruzual \& Charlot
(2003) code has been adopted.
The resulting
color at $z=0$ is typical of an S0 galaxy. Dotted and dashed curves are derived adopting
$\tau = 1$ Gyr and $\tau = 0.3$ Gyr, respectively, and could represent the reddest fraction of the sample.
}
\end{figure}

\begin{figure}
\plotone{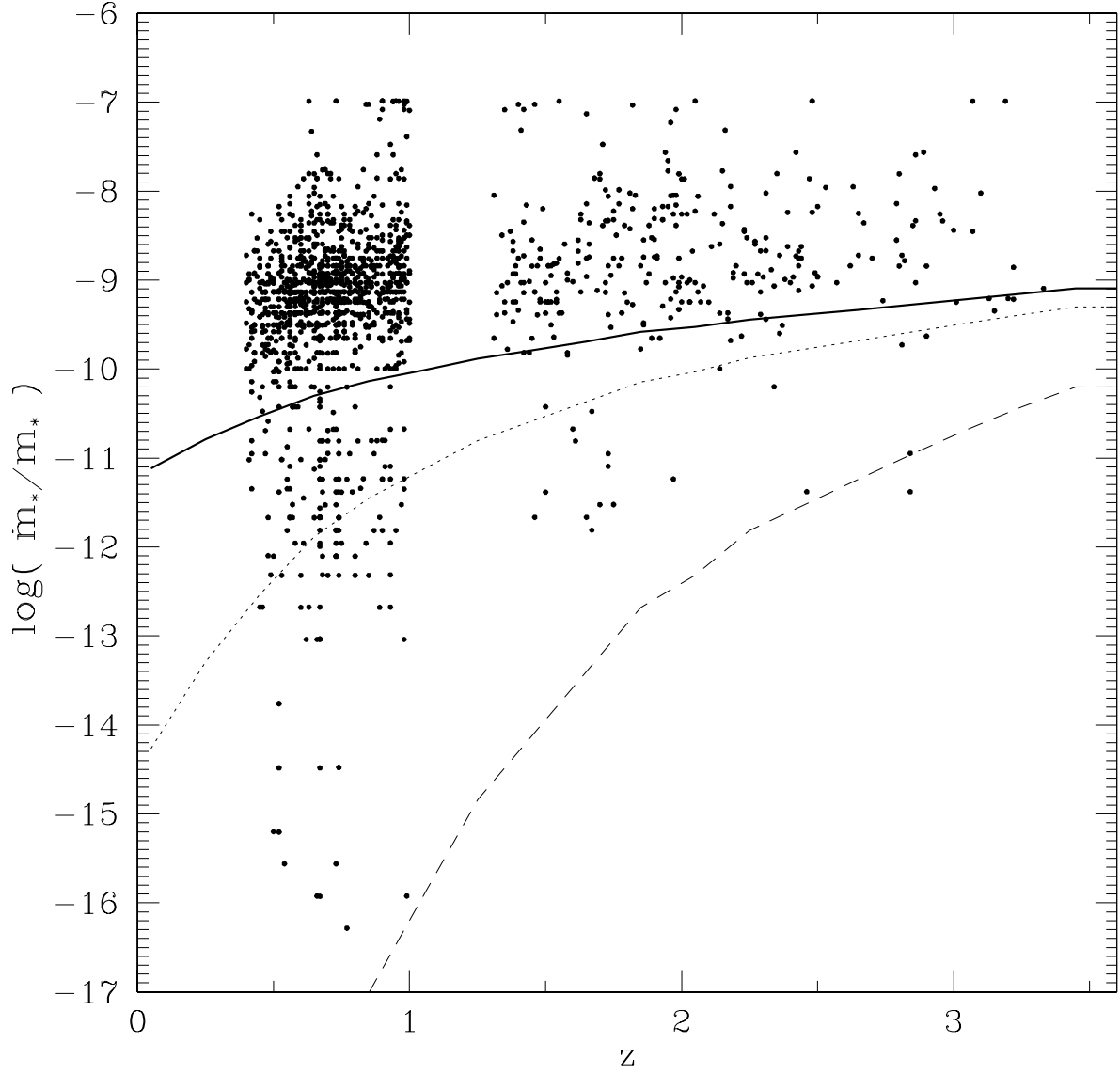}
\caption{Specific star formation rate $\dot m/m$ as a function of redshift.
The curves are obtained from the same models as in Fig.1}
\end{figure}

\begin{figure}
\plotone{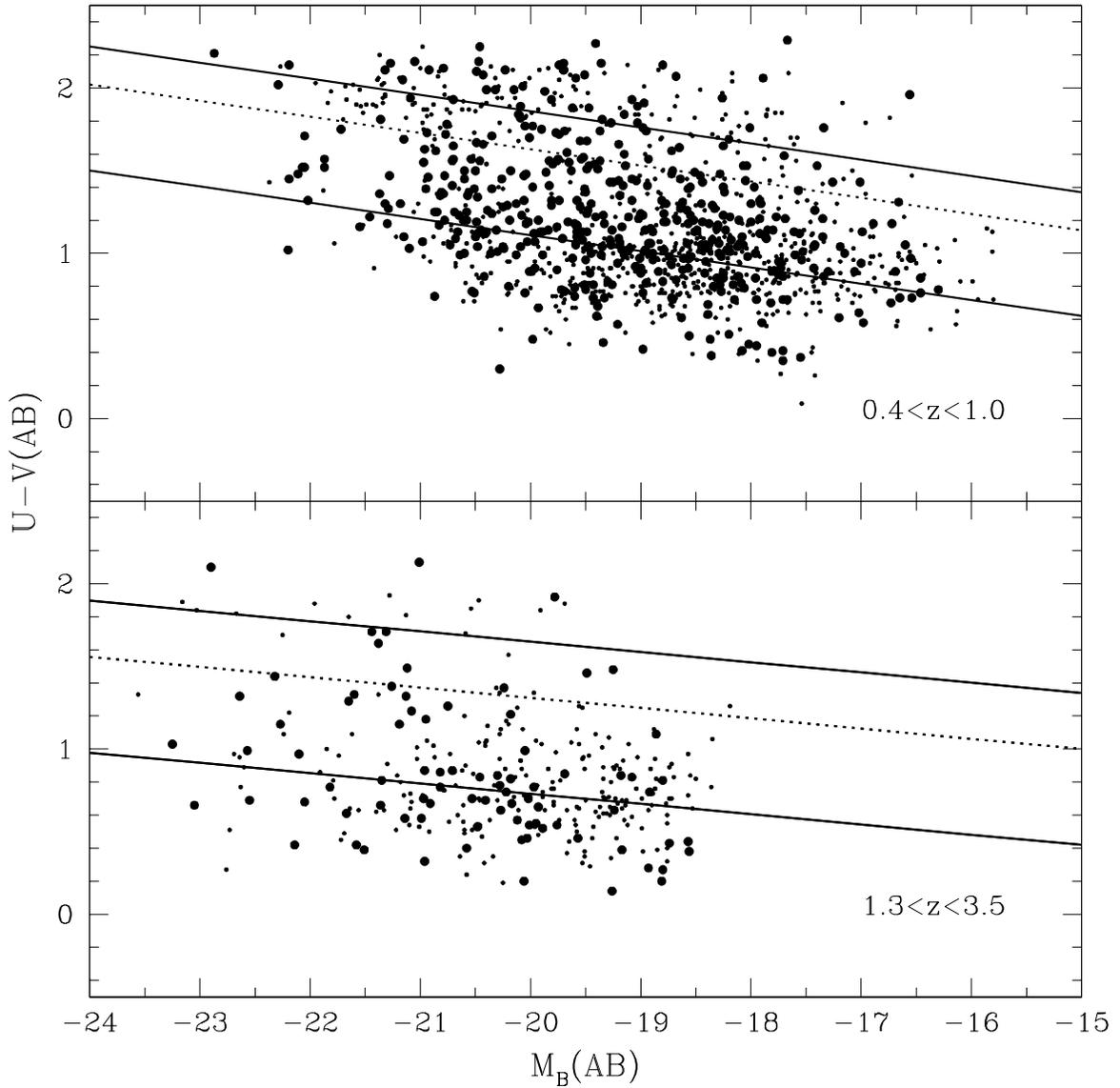}
\caption{Color magnitude relation in two redshift intervals. 
Black filled dots are galaxies with $E(B-V)\geq 0.2$. Small dots are galaxies
with $E(B-V)< 0.2$. Thick lines represent the best fit
relations for the red and blue populations. Dotted lines represent the locus of the minimum
between the two distributions used for the class separation.}
\end{figure}

\begin{figure}
\plotone{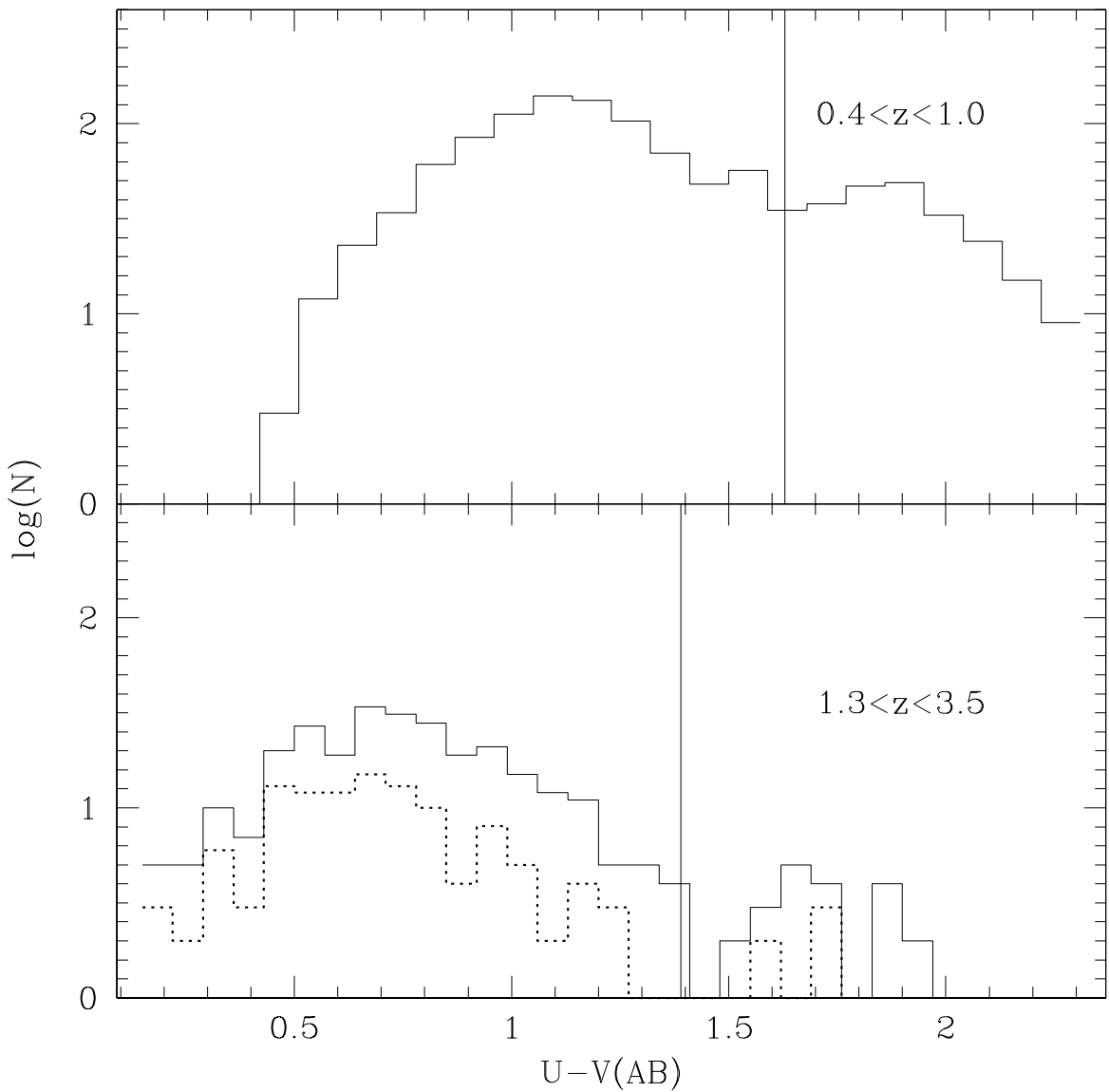}
\caption{Color histograms in two redshift intervals.  The continuous vertical lines represent the average
separation evaluated from the color bimodal distribution. The lines are evaluated
at the average redshift and at the average characteristic magnitude of the two Schechter populations ($M^*\simeq -19.9$
at $0.4<z<1$ and $M^*\simeq -21.2$ at $1.3<z<3.5$) for illustrative purpose. The dotted histogram 
represents the color distribution of galaxies in the redshift interval $2<z<3.5$.}
\end{figure}

\begin{figure}
\plotone{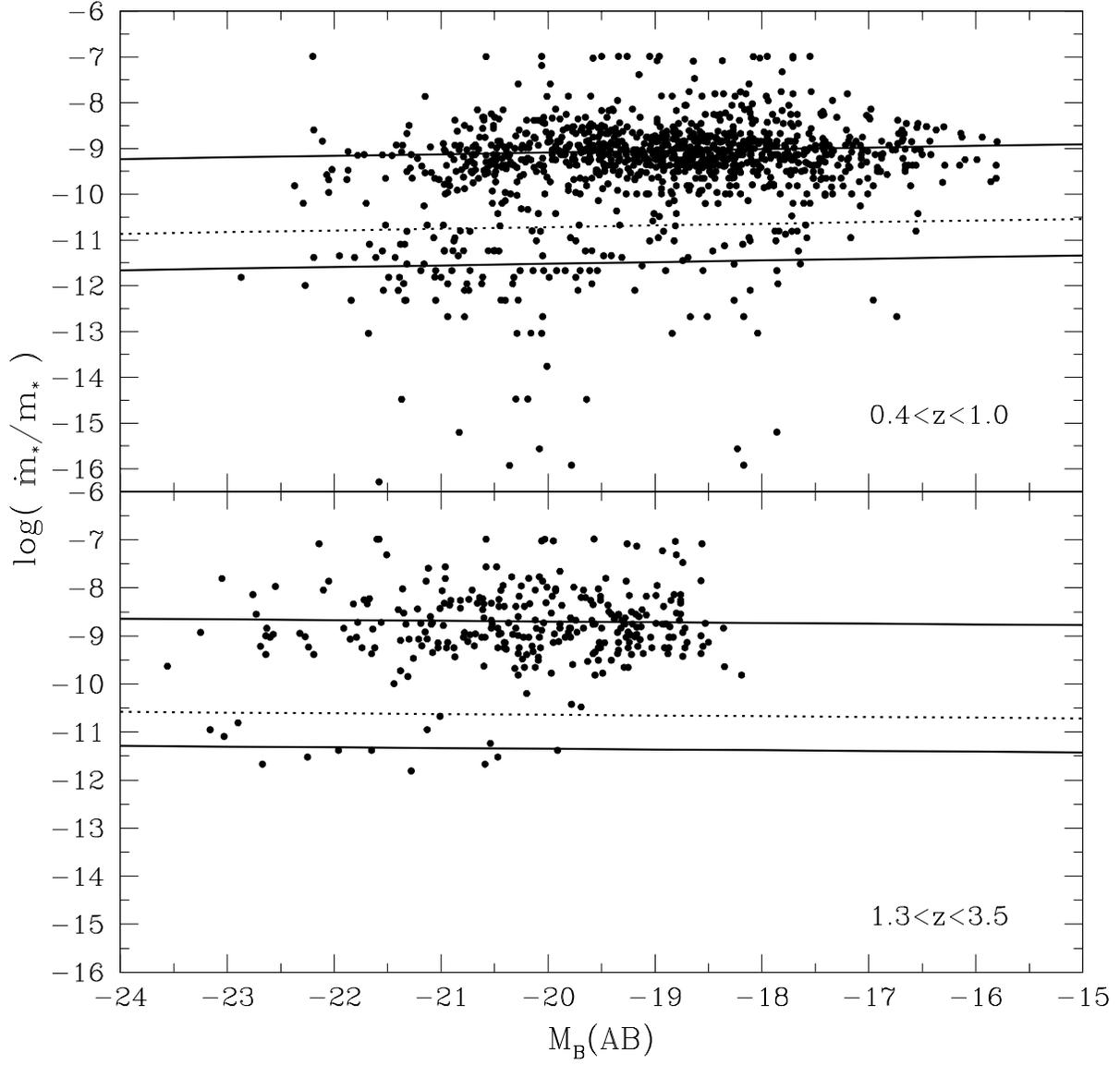}
\caption{Specific star formation rate $\dot m/m$ as a function of $M_B$ at different redshift intervals. 
Thick and dotted lines as in Fig. 3}
\end{figure}

\begin{figure}
\plotone{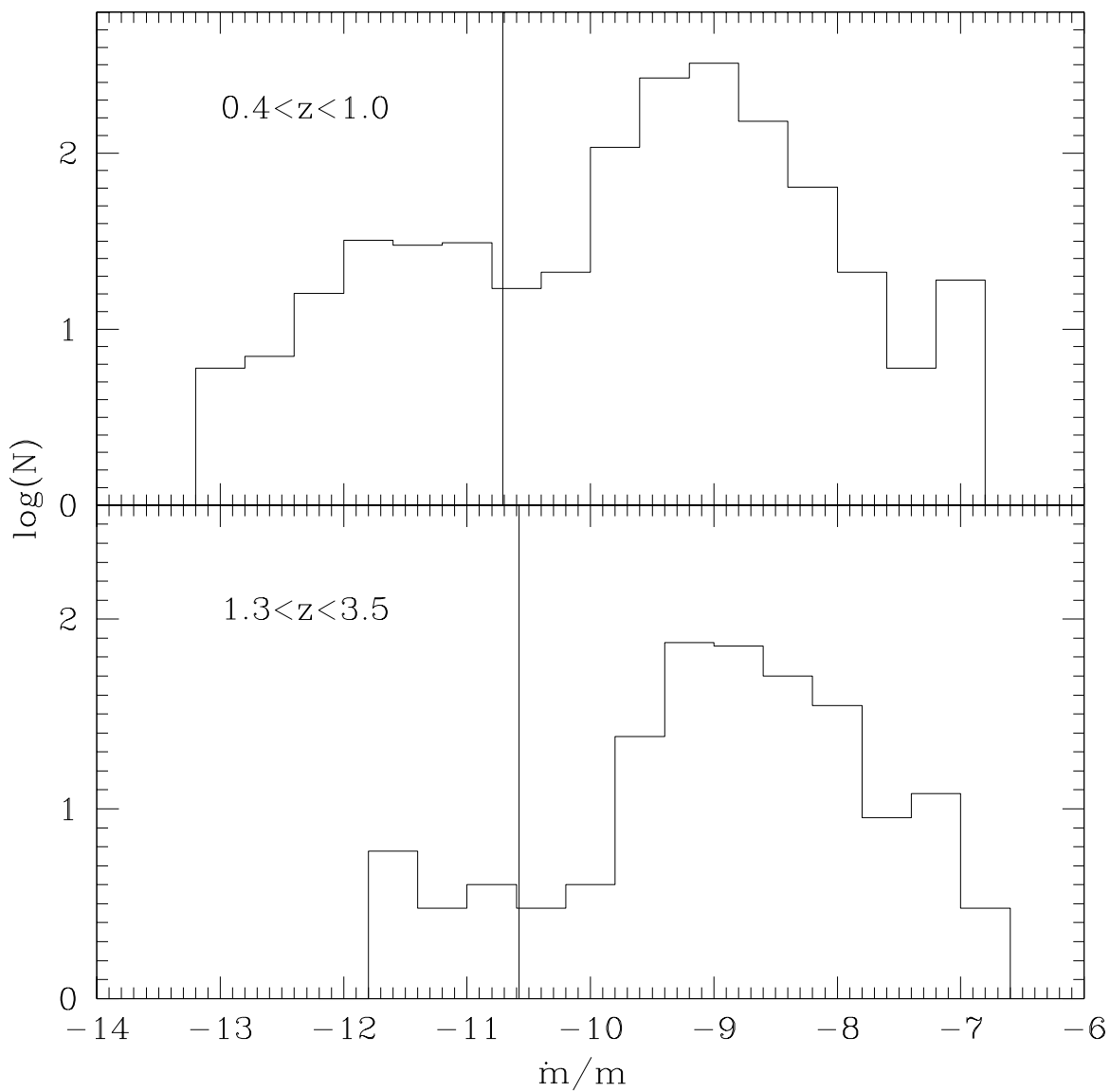}
\caption{SSFR histograms.  The continuous vertical lines represent the average
separation evaluated from the SSFR bimodal distribution. The lines are evaluated
at the average redshift and at the average characteristic magnitude of the two Schechter populations ($M^*\simeq -19.7$
at $0.4<z<1$ and $M^*\simeq -21.2$ at $1.3<z<3.5$) for illustrative purpose.}
\end{figure}

\begin{figure}
\plotone{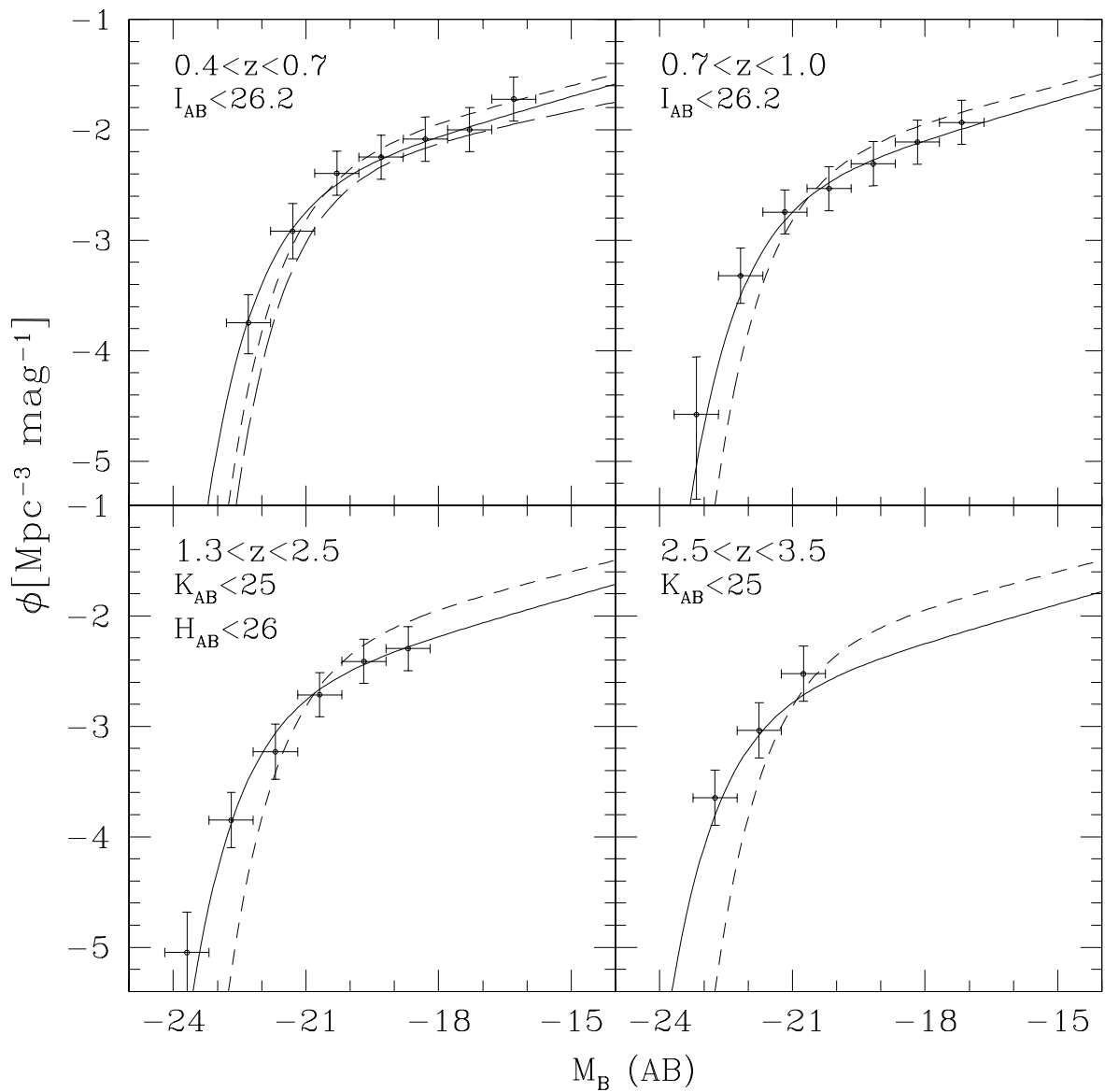}
\caption{Total luminosity function as a function of redshift. Continuous curves come from the 
Maximum Likelihood analysis.
The local LFs from the Sloan and 2dF surveys are also
shown as short and long-dashed curves, respectively, as in paper I. }
\end{figure}

\begin{figure}
\plotone{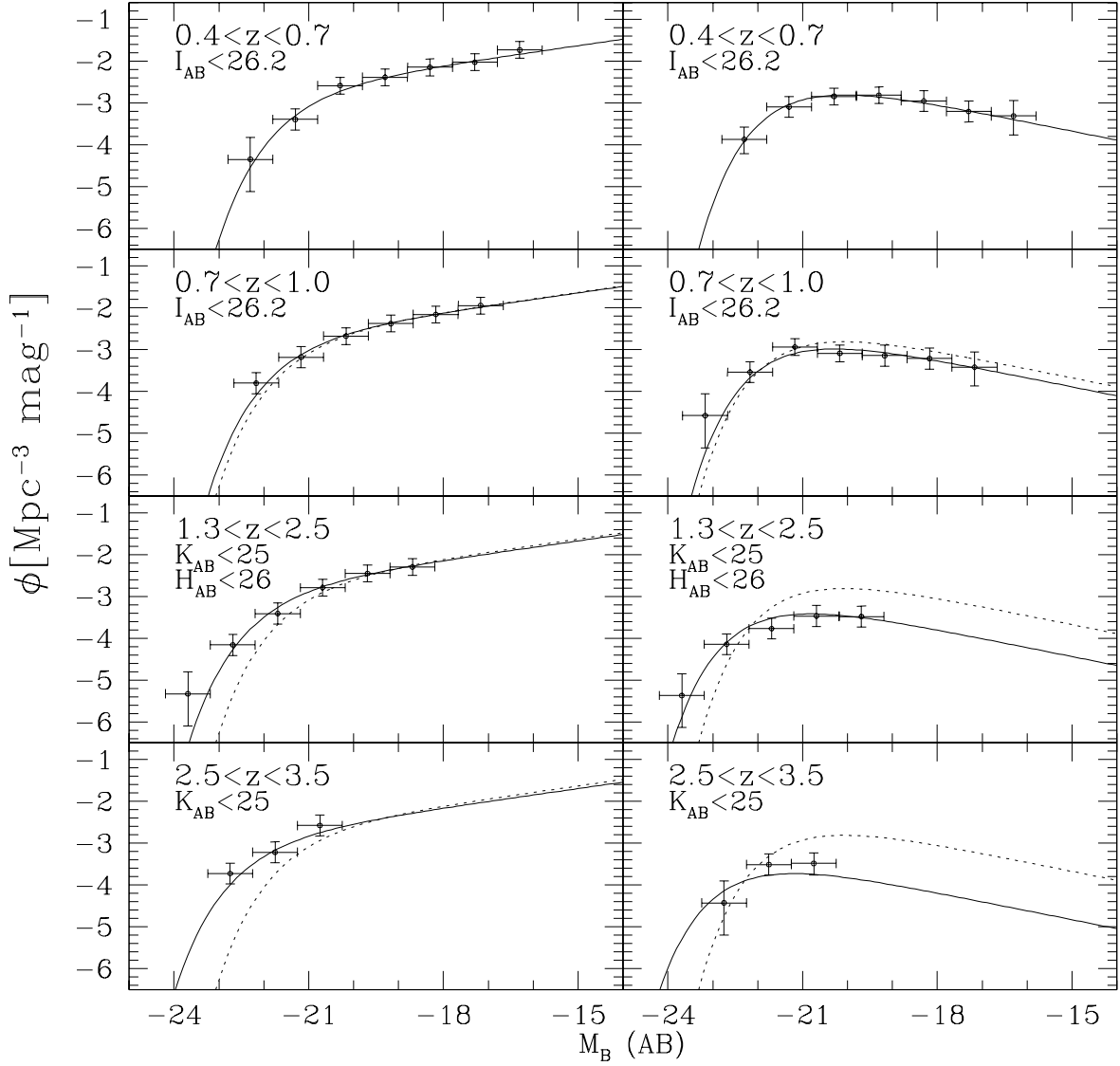}
\caption{Luminosity function of blue (left) and red (right) galaxies separated in the rest-frame 
$U-V$ color following the "S0 evolutionary
track " (see text for details).
Continuous curves come from the Maximum Likelihood analysis.
The luminosity function in the lowest redshift bin $0.4<z<0.7$ is also shown for comparison in all
panels (dotted curve).}
\end{figure}

\begin{figure}
\plotone{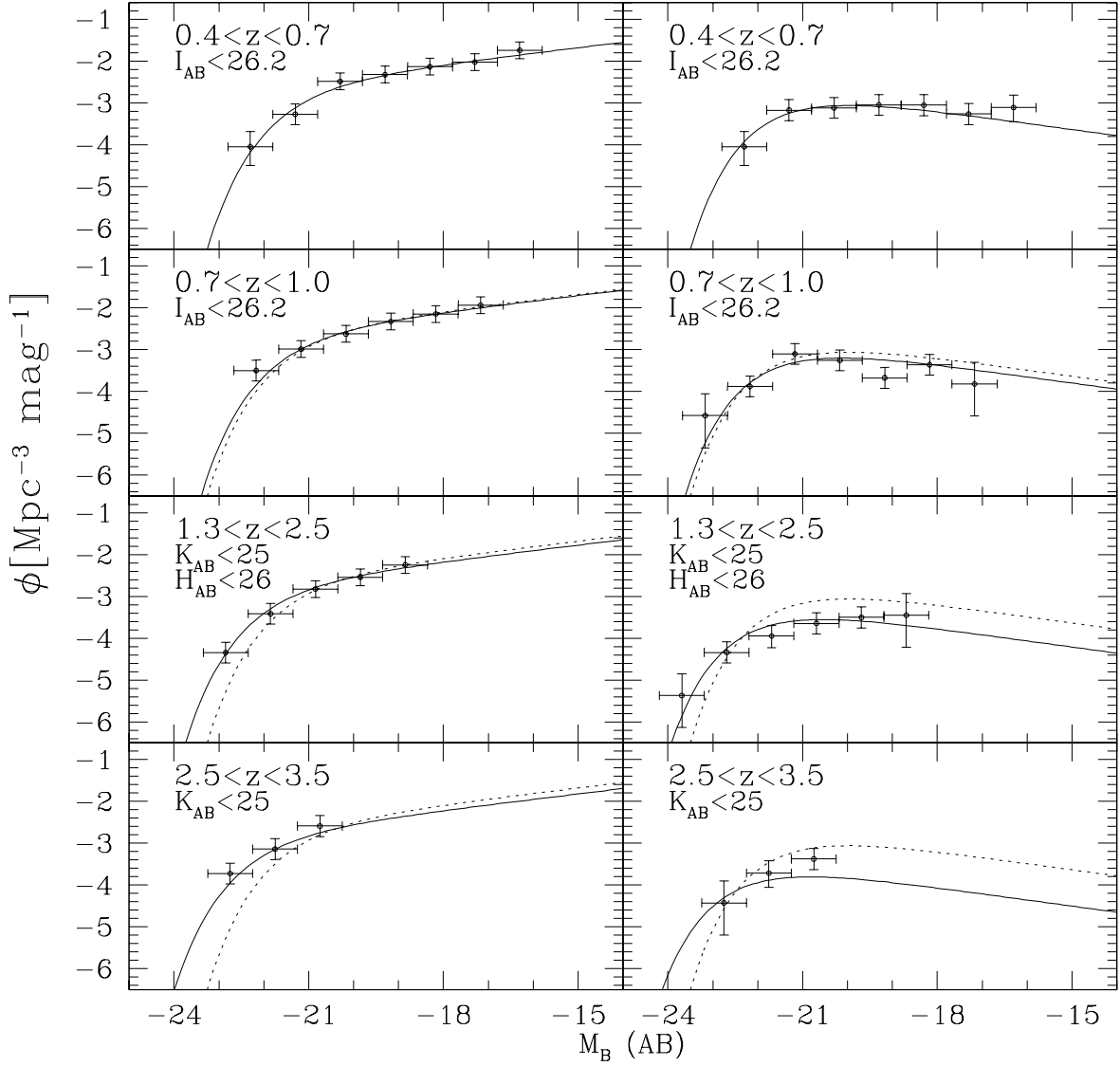}
\caption{Luminosity function of "late" (left) and "early" (right) galaxies
separated on the basis of their specific $\dot m_*/m_*$ 
star formation rate following the "S0 evolutionary track". 
The luminosity function in the lowest redshift bin $0.4<z<0.7$ is also shown for comparison in all
panels (dotted curve).
}
\end{figure}

\begin{figure}
\plotone{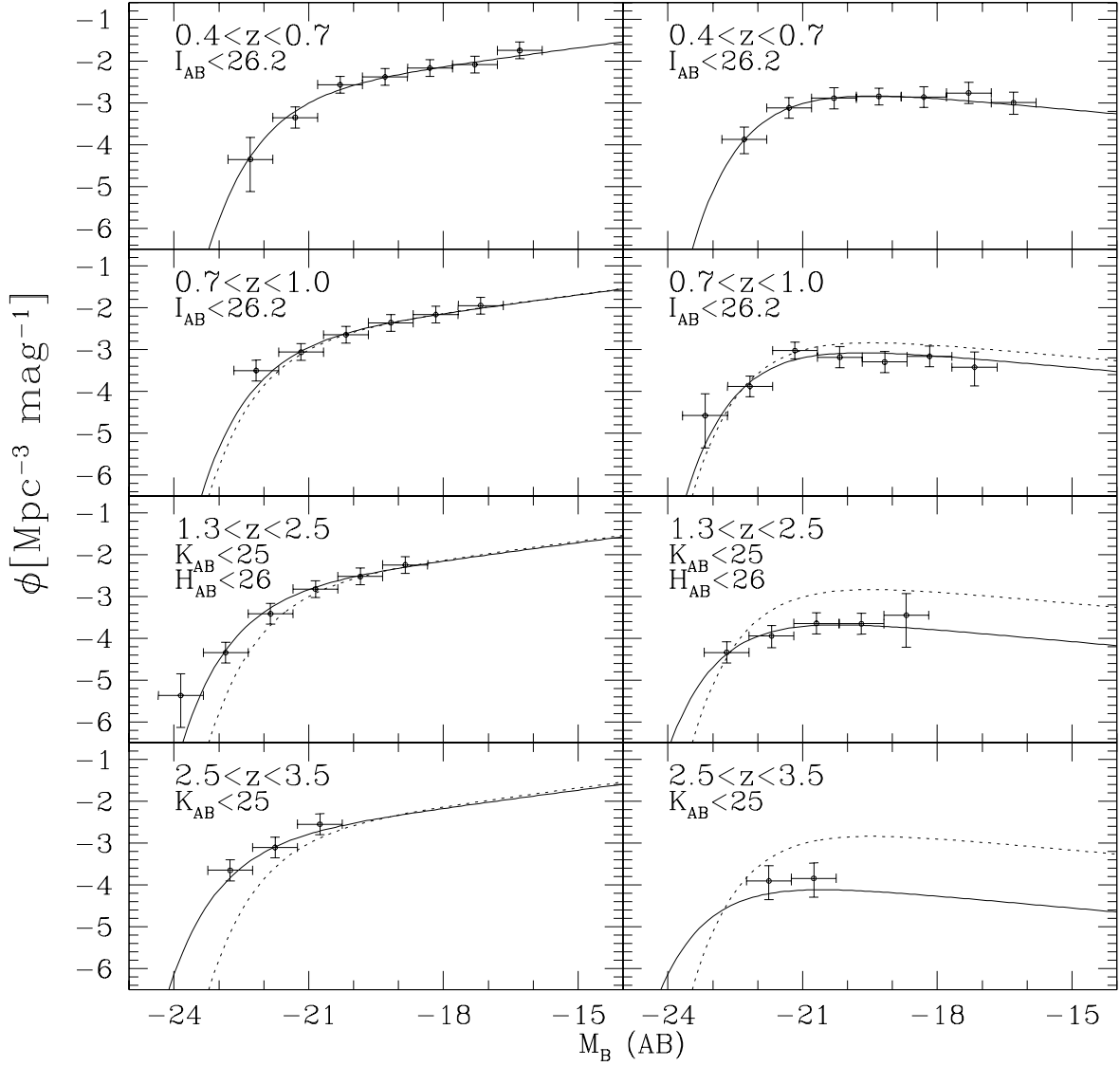}
\caption{Luminosity function of "late" (left) and "early" (right) galaxies
 separated following their rest-frame $U-V$
colors as derived from the minimum in their bimodal gaussian distributions. 
The luminosity function in the lowest redshift bin $0.4<z<0.7$ is also shown for comparison in all
panels (dotted curve).
}
\end{figure}

\begin{figure}
\plotone{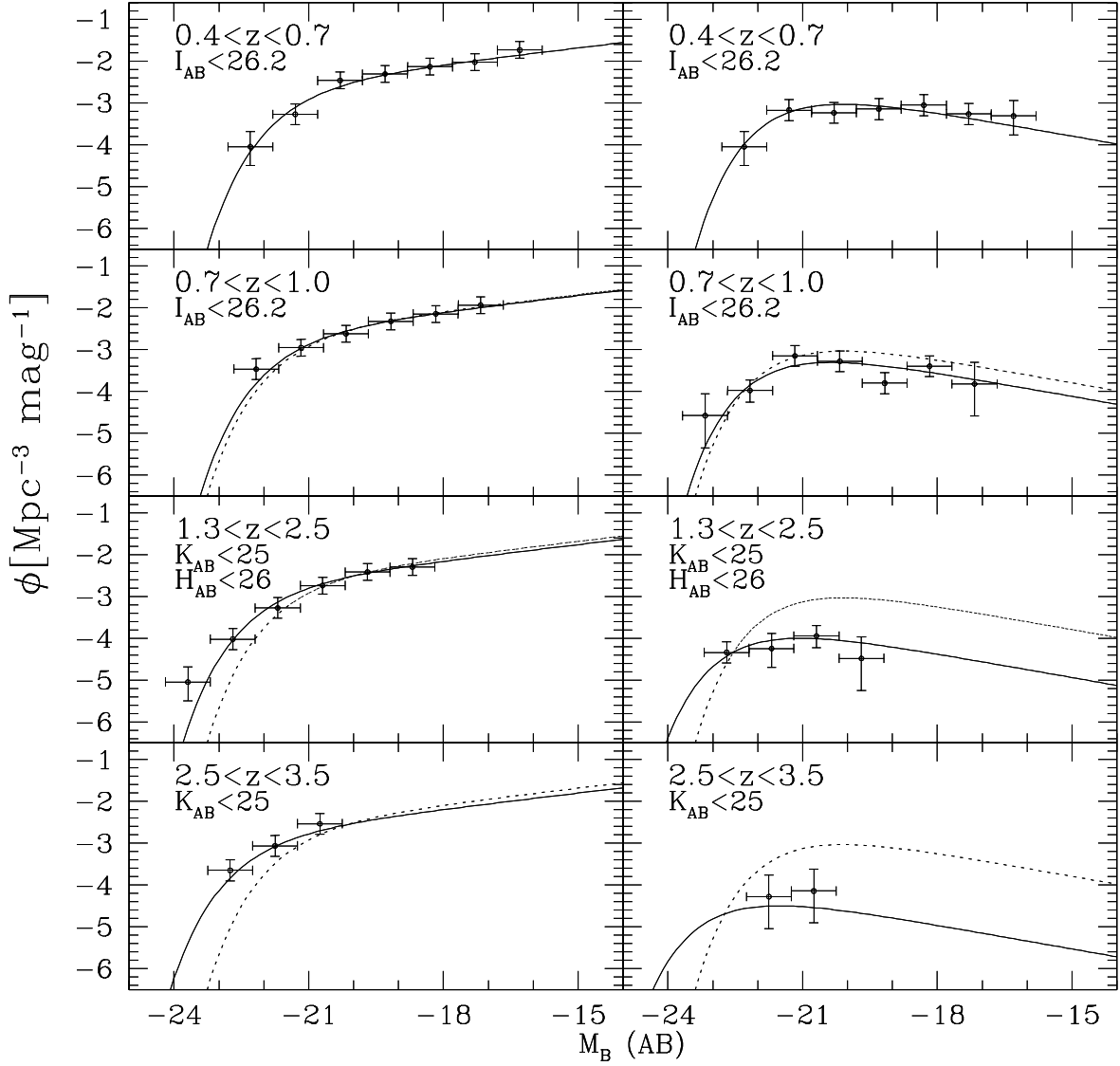}
\caption{Luminosity function of "late" (left) and "early" (right) galaxies separated following their specific $\dot m_*/m_*$ 
star formation rate as derived from the minimum in their bimodal gaussian distributions. 
The luminosity function in the lowest redshift bin $0.4<z<0.7$ is also shown for comparison in all panels
(dotted curve).
}
\end{figure}

\begin{figure}
\plotone{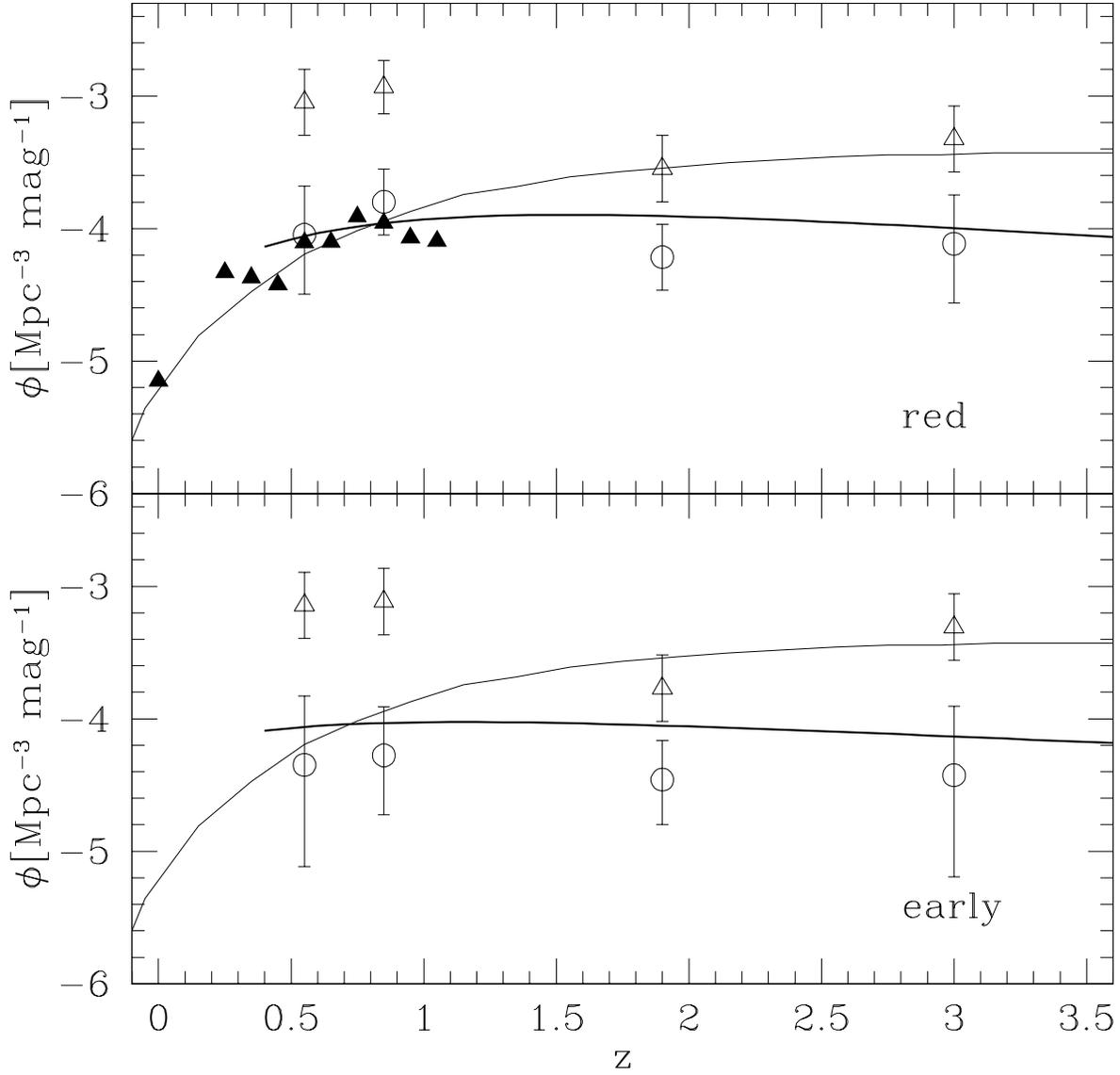}
\caption{Density evolution as a function of redshift for red/early type objects separated on the basis of
the "S0 evolutionary tracks".
Circles are galaxies in the range $-23<M_B<-22$ and in the same redshift intervals used to compute the LFs.
Open triangles are galaxies in the range $-21.5<M_B<-20.6$.
Filled triangles come from the Schechter LFs of Bell et al. (2004).
The thin continuous curve comes from the PLE model discussed in the text and computed at $M_B=-22.5$. 
The thick continuous curve comes from the integration of the best fit LFs in the range $-23<M_B<-22$.
Upper panel for the color selected
galaxies, lower panel for galaxies selected according to their specific star formation rate, as in Figs.7,8.}
\end{figure}

\begin{figure}
\plotone{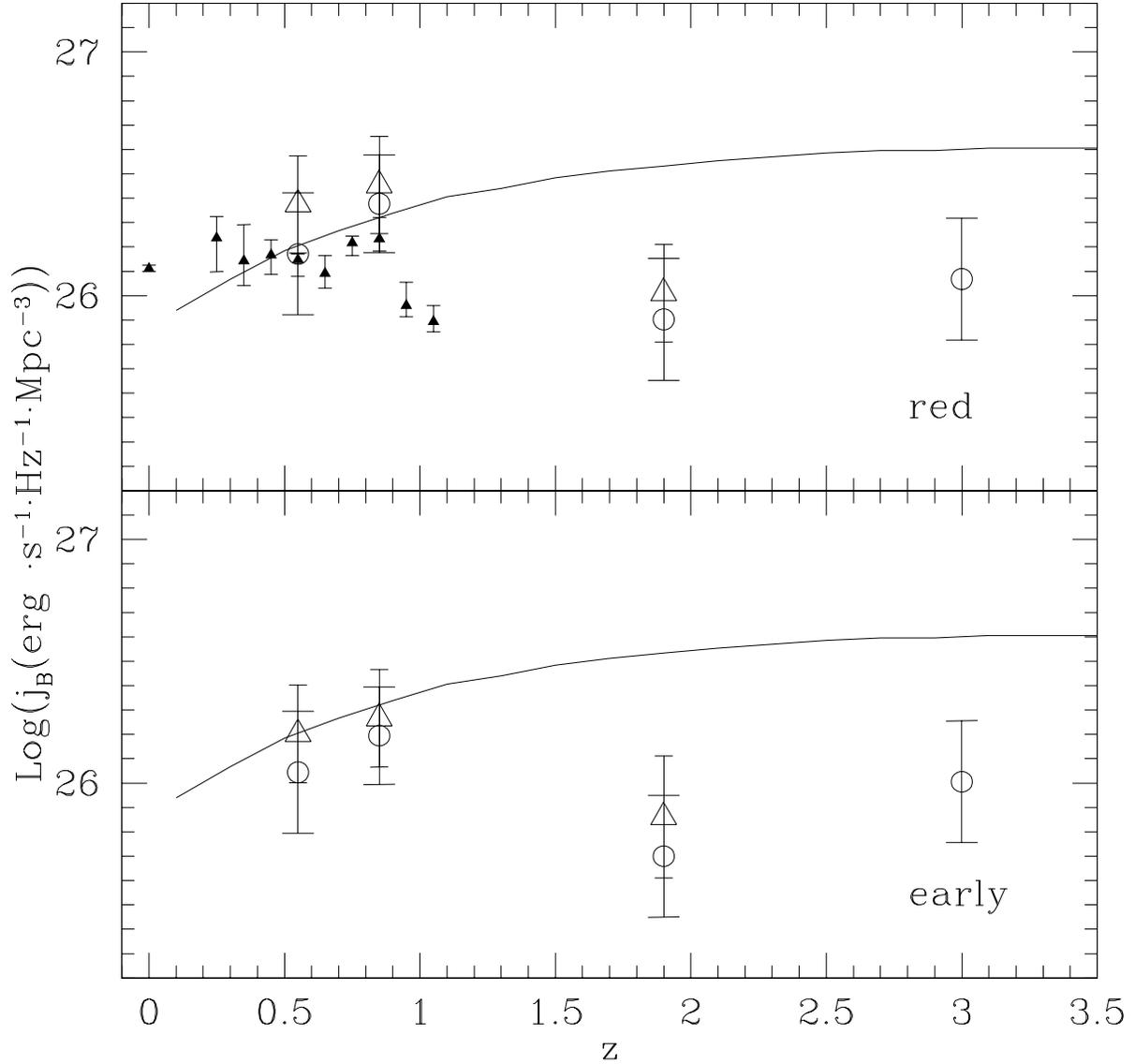}
\caption{Luminosity density in the rest-frame B-band of the red/early type galaxies selected as in Fig.12.
Galaxies separated by their rest-frame $U-V$ color
are shown in the upper panel, while those separated by their $\dot m_*/m_*$ are shown in the lower panel.
The open circles represent the contribution from bright galaxies with $M_B<-20.6$ which are at the faintest
limit of our sample at $z\sim 3$. Open triangles
represent the contribution from galaxies with $M_B<-19.5$ 
for comparison with the Bell et al. 2004 data (filled triangles) at $z<1$. The curve represents
the passive evolution model shown in the previous figure for galaxies with $M_B<-20.6$.}
\end{figure}

\begin{figure}
\plotone{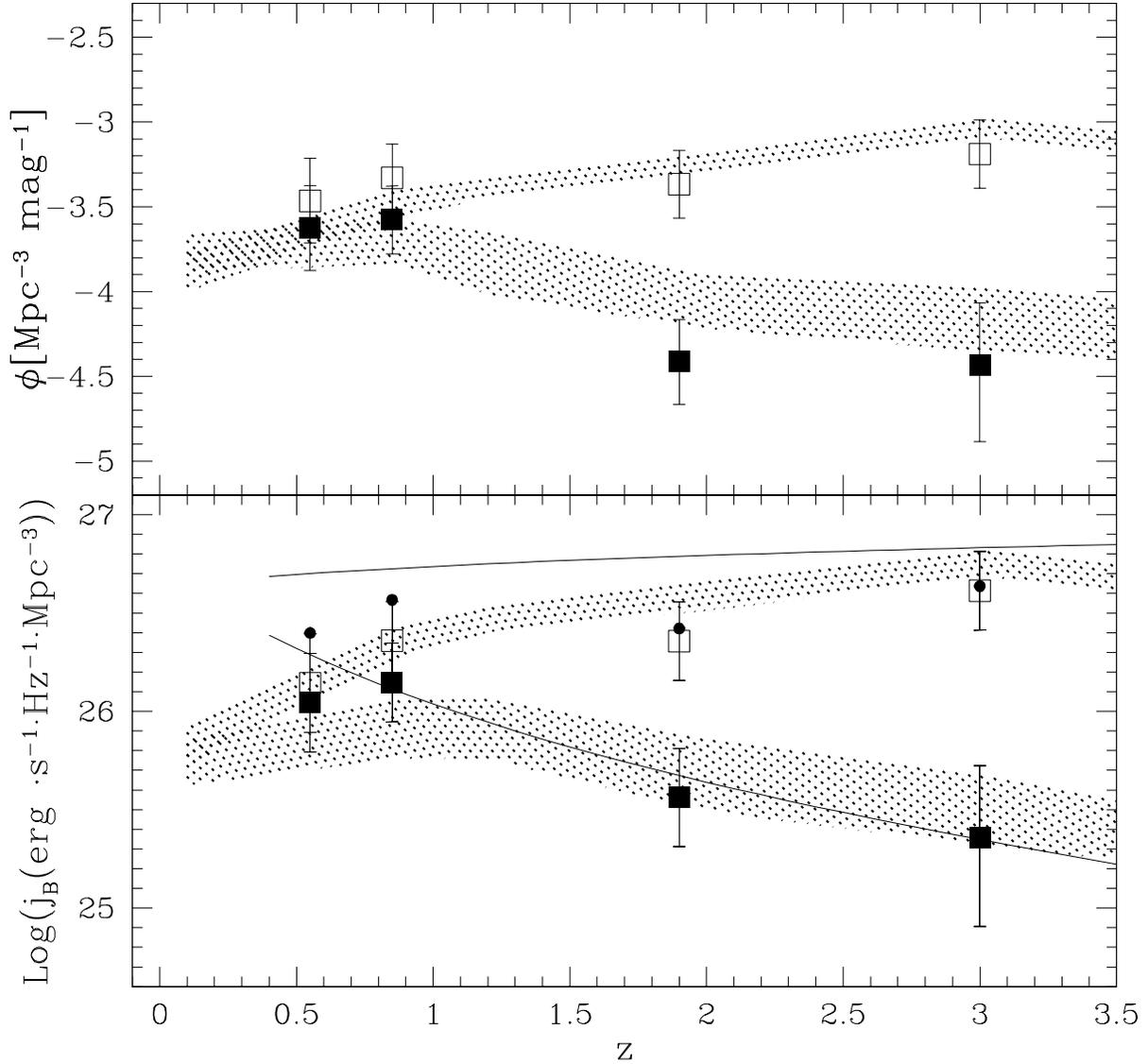}
\caption{Number and B-band luminosity densities of early (filled squares) and
late (open squares) type galaxies separated
by their bimodal $\dot m_*/m_*$ distribution. Only galaxies with $-24<M_B<-20.6$
are considered. The B-band luminosity density of the bright early plus late type galaxies is shown as
small filled circles in the lower panel. The continuous lines represent
the "total" luminosity densities of the early and late type galaxies obtained
after integration of the best fit LFs down to $M_B=-15.8$.
The shaded regions represent the uncertainties of the predictions
of our CDM model due to the spread in the $\dot m_*/m_*$ threshold adopted
and to the dust extinction curves (Small Magellanic Clouds, Milky Way and 
Calzetti) affecting the luminosity cut.}
\end{figure}

\end{document}